\documentclass[aps,prl,twocolumn,notitlepage,showpacs,preprintnumbers,amsmath,amssymb,revsymb]{revtex4-1}
\usepackage{graphicx}
\usepackage{color} 
\usepackage{txfonts}
\usepackage{dcolumn}
\usepackage{bm}
\usepackage{braket}
\usepackage[colorlinks=false, hidelinks]{hyperref}
\usepackage{pdfpages}
\makeatletter
\AtBeginDocument{\let\LS@rot\@undefined}
\makeatother

\definecolor{gray}{rgb}{0.5,0.5,0.5}

\def\braket#1{\mathinner{\langle{#1}\rangle}}

\begin{document}
\title{Robust quantum optimizer with full connectivity}
\author{Simon E. Nigg}\email[Corresponding author:
]{simon.nigg@unibas.ch}
\author{Niels L{\"o}rch}
\author{Rakesh P. Tiwari}
\date{\today}

\begin{abstract}
{Quantum phenomena have the potential to speed up the solution of hard
  optimization problems. For example quantum
annealing, based on the quantum tunneling effect, has recently been
shown to scale exponentially better with system size as
compared with classical simulated annealing. However, current realizations of
quantum annealers with superconducting qubits face two major challenges. First, the
connectivity between the qubits is limited, excluding many optimization
problems from a direct implementation. Second, decoherence degrades the success
probability of the optimization. We address both of these shortcomings
and propose an architecture in which the qubits are robustly encoded in continuous variable degrees of freedom. Remarkably, by leveraging the phenomenon of flux
quantization, all-to-all connectivity is obtained without
overhead. Furthermore, we demonstrate the robustness of this architecture by simulating the optimal
solution of a small instance of the NP-hard and fully connected number partitioning
problem in the presence of dissipation.}
\end{abstract}

\maketitle

Optimization problems are ubiquitous in nature and throughout human
activities such as computational biology, combinatorial chemistry or
corporate planing. Consider for example the following task: Given a set of
assets with fixed values, is it possible to share them fairly between
two parties? This decision problem is an instance of what is known as
the Number
Partitioning Problem (NPP), which is non-deterministic
polynomial-time hard (NP-hard)~\cite{GareyJohnson,Mertens-1998}. This
means that any known exact algorithm will take a time exponential in
the number of assets to
solve at least some instances of this problem. It
turns out that this problem, like many others~\cite{Lucas-2014},
is isomorphic to finding the ground state configuration of an anti-ferromagnetic
long range Ising model with free energy
\begin{align*}
E(s_1,s_2,\dots,s_N)=\sum_{ij}J_{ij}s_is_j,\quad J_{ij}>0,\quad s_i\in\{-1,1\}.
\end{align*}
Finding a ground state means finding a configuration
of spins $\{s_1,s_2,\dots, s_N\}$, which minimizes this energy. To map the NPP onto an Ising
problem for example, we set $J_{ij}\propto n_in_j$, where $n_i$ denotes the value of the $i$-th asset. The Ising
spin $s_i$ labels which of the two sets in a given partition of the
set of assets $A=F \cup\bar F$, the asset $i$ is in. If $s_i=+1$
then the asset is in $F$, while if $s_i=-1$ the asset is in the
complement $\bar F=A\backslash F$. A fair partition exists if and
only if the corresponding ground state energy is zero since then:
$\sum_{i\in F}n_i=\sum_{i\in\overline F}n_i$. If no
fair partition exists, minimizing the energy will
yield the least unfair partition.

It is noteworthy that for the NPP
approximate solutions obtained via classical heuristics,
such as simulated annealing~\cite{Kirkpatrick-1983}, can be very
poor~\cite{Hayes-2002}. In general, the optimization landscapes of hard problems are typically highly
non-convex and heuristic approaches tend to get trapped in local minima. In 1998, Kadowaki and Nishimori~\cite{Kadowaki-1998}, introduced
the idea that the phenomenon of quantum tunneling could help to
escape from local potential minima. This insight and subsequent
works~\cite{Brooke-1999, Lee-2000, Farhi-2001} have led
to high interest in quantum annealing both in academia and in the
private sector~\cite{Santoro-2002, Johnson-2011, Boixo-2014, Boixo-2016,
Neill-2016}. While a genuine quantum
speedup over the best known classical heuristic algorithms remains to be demonstrated, there are strong indications that
this could be achieved in the near future. In particular, for problem
instances with high and narrow barriers, the D-wave quantum annealer~\cite{Denchev-2016}
has been shown to succeed exponentially faster than thermally
activated simulated annealing.

\section*{Two main challenges}
Many hard optimization problems map onto Ising models with  dense
connectivity graphs. The D-wave architecture however has limited connectivity~\cite{Denchev-2016}. This is due to the fact that
the interconnections between qubits are realized by physical coupler
devices. Because the number of connections grows quadratically with
the number of qubits, such an approach quickly represents an
intractable design challenge. 
To circumvent this connectivity
problem,
the standard approach is to use a minor embedding (ME) scheme~\cite{Choi-2008,Choi-2011} to map a fully connected graph onto the
sparse physical graph. However, this comes at the cost
of a
substantial overhead in the number of physical qubits. More recently another mapping was proposed, where
instead of the $N$ bits of the Ising model, the
$N(N-1)/2$ binary pairings are encoded in a two-dimensional lattice of
qubits with four-qubit nearest neighbor interactions~\cite{Lechner-2015}. Implementations of this idea with superconducting
circuits have been proposed in Refs.~\cite{Leib-arxiv-2016} and~\cite{Chancellor-arxiv-2016}. Also in this case there remains a substantial overhead as
compared to a direct implementation. In contrast, here we propose a superconducting
continuous variable Ising machine (CVIM) with full connectivity and zero
overhead, paving the way to experimentally demonstrate quantum supremacy~\cite{Boixo-arxiv-2016}.

Ideally, quantum annealing relies on the coherent evolution of
the ground state of an isolated quantum system, the Hamiltonian of
which is varied in time adiabatically~\cite{Born-1928}. Typically, the Ising
spin $s_i$ is encoded in the two states of a qubit and the Hamiltonian is
of the form (we set $\hbar=1$)
\begin{align}\label{eq:12}
\bm H_A(t)=\varepsilon(t)\sum_{j}\bm\sigma_j^x+\bm H_{\rm Ising}.
\end{align}
The value of the control parameter $\varepsilon(t)$ is initially
chosen such that $|\varepsilon(t=0)|\gg \max|J_{ij}|$ and the system
is prepared in the ground state of $\bm H_A(0)$. $\varepsilon$ is then gradually
reduced to zero such that $\bm H_A(T)=\bm H_{\rm
  Ising}=\sum_{ij}J_{ij}\bm\sigma_i^z\bm\sigma_j^z$. If the adiabatic
condition is satisfied~\cite{Born-1928}, the system at time $T$ is in the ground state
of the quantum Ising model. Real world quantum
annealers however are open quantum systems. It was recognized
early on that one of the potential strengths of quantum annealing is
its relative robustness to certain types of errors affecting the underlying
qubits~\cite{Childs-2001, Sarandy-2005, Aberg-2005, Roland-2005,Tameem-2015}. More precisely, if decoherence takes place only in the
instantaneous energy eigenbasis of~(\ref{eq:12}), then it does not decrease the success
probability of the optimization as compared with the coherent
limit~\cite{Childs-2001}. In general however, decoherence takes place in
different channels depending on the
physical hardware. Dephasing errors in the Ising basis for example,
described by the random action of $\bm\sigma_j^z$ operators, are
deleterious to the success of the optimization, because the error
operators do not commute with the transverse part of the
Hamiltonian~(\ref{eq:12}) and hence lead to transitions out of the
ground state in the initial stages of the
annealing process~\cite{Boixo-2016}. In contrast, here we show that quantum
annealing with our continuous variable system is remarkably robust to decoherence. A key result of the present work is that if the annealing rate is below
the problem specific adiabatic threshold but large
compared with the dissipation rate, then a single run of the
continuous variable optimizer will succeed with
probability $>0.5$ for ramp rates smaller than but arbitrarily close to the
threshold value.

\section*{Continuous variable Ising machine}

The binary
Ising spin variable $s_i$ is encoded into the quantized phase of a
Kerr parametric oscillator~\cite{Yamamoto-2008,Wustmann-2013,supp_mat,Goto-2016,
  Goto-2016b, Puri-arxiv-2016}
(KPO) above threshold, which can take on two values: $0$,
corresponding to $s_i=+1$ or $\pi$,  corresponding
to $s_i=-1$. The
dynamics of such a system is
described by the Hamiltonian
\begin{align*}
\bm H_{\bm a}(\Delta, K)=\Delta\bm a^{\dagger}\bm a+\varepsilon\left( \bm
  a^2+\bm a^{\dagger 2} \right)-K\bm a^{\dagger}\bm a^{\dagger}\bm
  a\bm a.
\end{align*}
\begin{figure}[ht]
\includegraphics[width=\columnwidth]{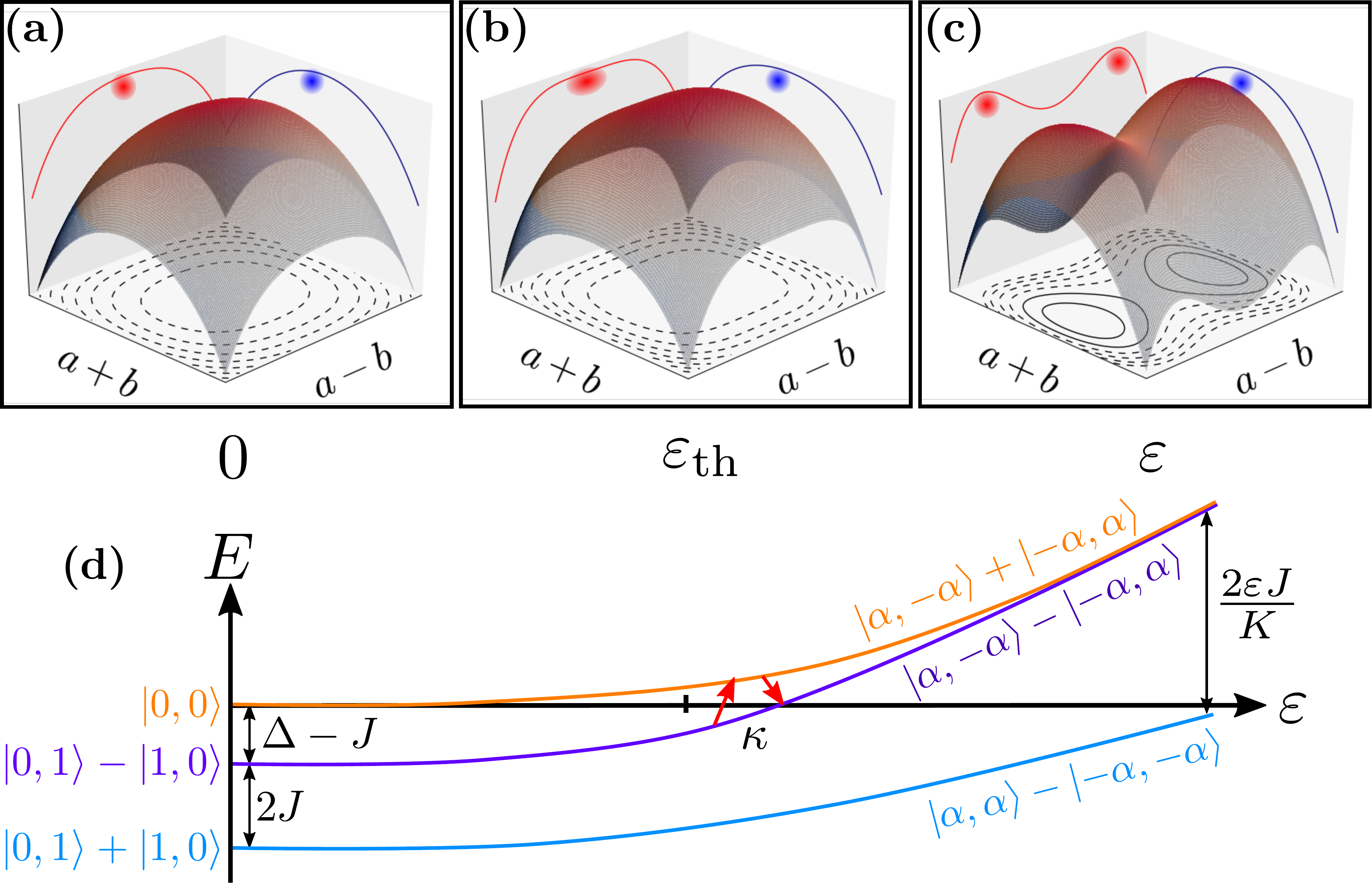}\caption{Illustration
  of the bifurcation-based annealing process in two anti-ferromagnetically coupled KPOs. Panels {\bf (a)}, {\bf (b)} and {\bf (c)} show the
  evolution of the 2D energy landscape with increasing two-photon drive
  strength $\varepsilon$. Shown is the antiferromagnetic case ($J<0$), where the
  anti-symmetric mode $\bm a-\bm b$ is softer, and
  thus has a lower bifurcation threshold, than the symmetric
  mode $\bm a+\bm b$. Consequently, the
  system evolves from the vacuum $\ket{0,0}$ at $\varepsilon=0$ to
  the two-mode cat state
  $\ket{\phi_+}=(\ket{\alpha,-\alpha}+\ket{-\alpha,\alpha})/\sqrt{2}$
  at large $\varepsilon$, as shown in panel {\bf (d)}. Photon loss events predominantly take
  place above threshold and induce transitions between the cat states
  $\ket{\phi_+}$ and
  $\ket{\phi_-}=(\ket{\alpha,-\alpha}-\ket{-\alpha,\alpha})/\sqrt{2}$,
  as indicated by the red arrows in panel {\bf (d)}.\label{fig:annealing}}
\end{figure}
Here $\bm a$ and $\bm a^{\dagger}$ are bosonic annihilation and
creation operators, $K>0$ is the strength of the Kerr nonlinearity and
$\varepsilon$ is the strength of a two-photon drive. If $\Delta <0$, then at $\varepsilon=0$ the
vacuum is the ground state of the system. As the drive strength is
increased, the system undergoes a bifurcation at the
threshold value $\varepsilon_{\rm th}=|\Delta|$ into
a superposition of coherent states
$(\ket{\alpha}+\ket{-\alpha})/\sqrt{2}$ with $\alpha\simeq\sqrt{\varepsilon/K}$, also called a cat state.

In Ref.~\cite{Goto-2016, Goto-2016b}, a system of $N$ KPOs,
coupled via a term of the form $\sum_{nm}J_{nm}\bm a_n^{\dagger}\bm
a_m^{}$ was considered. It was shown, using perturbation theory, that as the two photon drive
strength of each KPO is varied from $\varepsilon=0$ to
$\varepsilon\gg|\Delta|$, the multi-mode vacuum $\ket{0_1,0_2,\dots, 0_N}$ is adiabatically
connected to a multi-mode cat state of the form
\begin{align}\label{eq:1}
\frac{1}{\sqrt{2}}\left( \ket{s_1\alpha,s_2\alpha,\dots,s_N\alpha}+\ket{-s_1\alpha,-s_2\alpha,\dots,-s_N\alpha} \right),
\end{align}
where $s_1,\dots, s_N\in\{-1,1\}$ are such that the Ising energy
$-\sum_{n,m}J_{nm}s_ns_m$ is minimized. Such a system thus presents
the opportunity to encode an Ising optimization problem in the
adiabatic dynamics of a continuous variable quantum system. To build
some intuition we consider first the simple case of two identical coupled
KPOs and denote their operators with $\bm a$ and $\bm b$. The mutual coupling
then has the form $J(\bm a^{\dagger}\bm b+\bm
b^{\dagger}\bm a)$. In the
weakly nonlinear limit $K\ll 4|J|$, the Hamiltonian of this system can be
conveniently written in the basis of the symmetric $\bm d=(\bm a+\bm
b)/\sqrt{2}$ and anti-symmetric $\bm c=(\bm a-\bm b)/\sqrt{2}$ modes
as (see Methods)
\begin{align*}
\bm H_2=\bm H_{\bm d}(\Delta+J,K/2)+\bm H_{\bm c}(\Delta-J,K/2)-K\bm
  d^{\dagger}\bm d\bm c^{\dagger}\bm c.
\end{align*}
This corresponds to two KPOs with {\em different} frequencies, coupled via a {\em cross}-Kerr term
$-K\bm d^{\dagger}\bm d\bm c^{\dagger}\bm c$. If $\Delta\pm J<0$,
then the two-mode vacuum $\ket{0,0}$ is the ground state. Because of
the different frequencies the thresholds of the two modes are
shifted to $|\Delta+J|$ for the symmetric mode and to $|\Delta -J|$ for the
anti-symmetric mode. Hence, as the two-photon drive strength of the
two KPOs is increased, the soft mode ($\bm d$ if $J>0$ and $\bm c$ if $J<0$), undergoes a
bifurcation before the hard mode ($\bm c$ if $J>0$ and $\bm d$ if
$J<0$). Importantly, once the soft mode starts to bifurcate and
becomes populated with photons, the cross-Kerr coupling makes the
hard mode even harder, pushing its bifurcation threshold further
away. Hence, the Kerr nonlinearity provides a stabilizing
feedback mechanism. This is schematically illustrated in Fig.~\ref{fig:annealing}~(a-c).

\section*{Physical implementation}
We next
turn to the physical implementation of such a machine with
superconducting circuits. Prototypes of similar bifurcation based coherent Ising machines have been built
with optical systems~\cite{Wang-2013, Hamerly-2015, Takata-2015}. Up until now however, no physical realization has
been proposed that implements a dense connectivity graph in a scalable
fashion. As compared with optical systems, superconducting Josephson circuits
offer the crucial advantage of stronger nonlinearities that have
recently reached the quantum regime, where the nonlinear frequency
shifts are larger than the resonance linewidths~\cite{Nigg-2012,
  Bourassa-2012, Kirchmair-2013}.

\begin{figure}[ht]
\includegraphics[width=0.9\columnwidth]{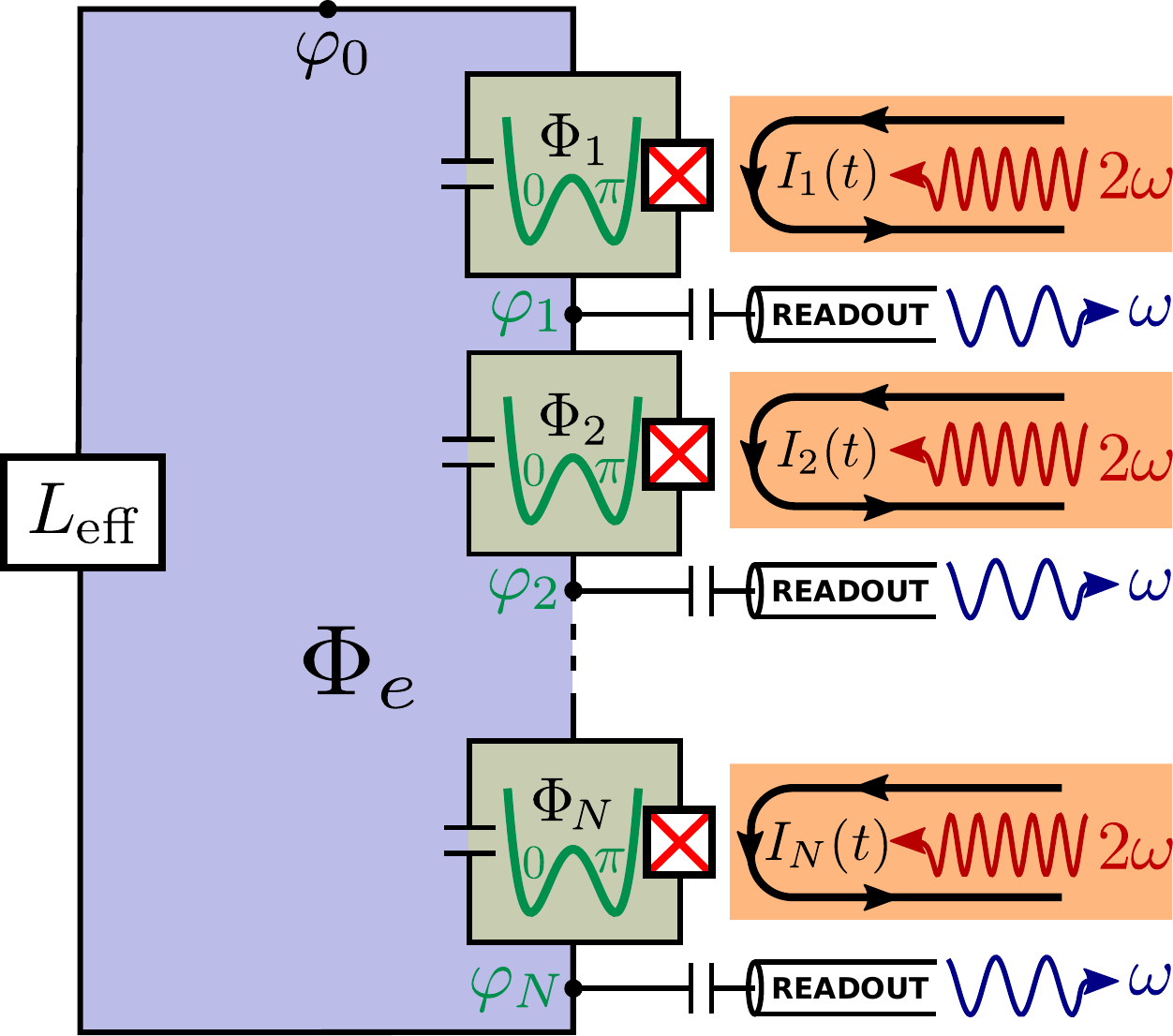}
\caption{Schematics of the proposed superconducting continuous
  variable Ising machine (CVIM). It
  consists of a chain of split Josephson junctions shunted by an
  effective inductor $L_{\rm eff}$. An ac modulation of the flux bias
through the split junction loops at twice their resonance frequency turns them into
KPOs with a pitchfork bifurcation (indicated by the
double well potentials). The Ising spin variables are encoded in
the quantized oscillation phases (either $0$ or $\pi$) of these
oscillators above threshold. An effective inductive shunt induces
all-to-all coupling between the KPOs. In order to obtain
anti-ferromagnetic coupling, a large area Josephson junction can be
used as a shunt together with a flux bias of $\Phi_e=\Phi_0/2$. Homodyne readout of the
oscillator phases is enabled via capacitively coupled transmission
lines~\cite{supp_mat}.}\label{fig:setup}
\end{figure}

A single KPO can be engineered with superconducting circuits by modulating the flux through a split
Josephson junction (Fig.~\ref{fig:setup}) at close to twice its natural resonance
frequency~\cite{Puri-arxiv-2016, supp_mat}. The Kerr nonlinearity is
provided by the Josephson potential expanded to fourth order in a
regime where the Josephson energy exceeds the charging energy.

While several superconducting KPOs can be coupled inductively, this typically yields
only a short range interaction between nearest neighbors. Long range
coupling between all pairs of $N$ oscillators can however be obtained
if the oscillators are connected in series and shunted by an inductive
element as shown in Fig.~\ref{fig:setup}. Indeed, flux quantization imposes a
constraint on the sum of all phase drops across the KPOs
($\{\varphi_n\}_{n=1,\dots, N}$) and across
the shunt ($\varphi_0$) such that
$\varphi_0=\sum_{n=1}^N\varphi_n$. The inductive energy of the shunt
\begin{align*}
\frac{\varphi_0^2}{2L_{\rm eff}}=\frac{1}{2L_{\rm eff}}\left( \sum_{n=1}^N\varphi_n^2+\sum_{n\not=m}\varphi_n\varphi_m \right),
\end{align*}
then immediately yields an
interaction term between all KPOs.
After including the capacitive energies and proceeding with standard circuit
quantization~\cite{Devoret-1995a}, $\varphi_n\rightarrow \sqrt{Z_n/2}(\bm a_n^{}+\bm
a_n^{\dagger})$, where $Z_n$ denotes the mode impedance, we obtain~\cite{supp_mat} the
effective Hamiltonian in a frame rotating with half the ac flux modulation
frequencies $\Omega_n=2(\omega_n-\Delta)$
\begin{align}\label{eq:4}
\bm H_{N}&=\Delta\sum_{n=1}^N\bm a_n^{\dagger}\bm
  a_n^{}+\sum_{n=1}^N\varepsilon_n\left(\bm
  a_n^2+\bm a_n^{\dagger 2} \right)\nonumber\\
&+\sum_{n\not=m}J_{nm}e^{i(\omega_n-\omega_m)t}\bm a_n^{\dagger}\bm a_m^{}-\sum_{n=1}^NK_n\bm
  a_n^{\dagger}\bm a_n^{\dagger}\bm a_n^{}\bm a_n^{}.
\end{align}
Here $\omega_n\simeq 4\sqrt{E_C^{(n)}E_J^{(n)}\cos(\Phi_n^{\rm
    dc})}$ is the natural frequency of the $n$-th KPO expressed in
terms of the charging and Josephson energies~\cite{Devoret-1995a} of the Josephson
junctions $E_C^{(n)}$ and $E_J^{(n)}$ and the dc part of the flux bias
$\Phi_n^{\rm dc}$. The two photon drive strength is
$\varepsilon_n\simeq (E_J^{(n)}/4)Z_n\sin(\Phi_n^{\rm
  dc})\delta\Phi_n^{\rm ac}$, where $Z_n=8E_C^{(n)}/\omega_n$ and
$\delta\Phi_n^{\rm ac}$ is the strength of the ac flux modulation at
frequency $\Omega_n$. We have assumed $|\Delta|\ll\omega_n$ and
applied the rotating wave approximation to suppress fast rotating terms. In the regime $E_J^{(n)}\gg
E_C^{(n)}$, the strength of the Kerr nonlinearity of
oscillator $n$ is
$K_{n}\simeq E_C^{(n)}/2$ and finally, the interaction
strength between oscillators $n$ and $m$ is $J_{nm}\simeq
\sqrt{Z_nZ_m}/(2L_{\rm eff})$.
In order to achieve all-to-all coupling, the $N$ KPOs must be made
resonant with each other. This can be achieved by tuning the $N$ dc
flux biases $\Phi_n^{\rm dc}$ such that
$\omega_1\simeq\omega_2\simeq\dots\simeq\omega_N$. Note that if the
shunt is a conventional inductor, the
coupling strengths so far are all positive, implying that only
ferromagnetic instances of the Ising model can be accessed. While some
non-trivial optimization problems can be mapped onto ferromagnetic
Ising models with inhomogeneous longitudinal fields~\cite{Boixo-2016},
others such as the NPP require anti-ferromagnetic couplings. Moreover,
anti-ferromagnetic couplings can give rise to
\begin{figure*}[ht]
\includegraphics[width=\columnwidth]{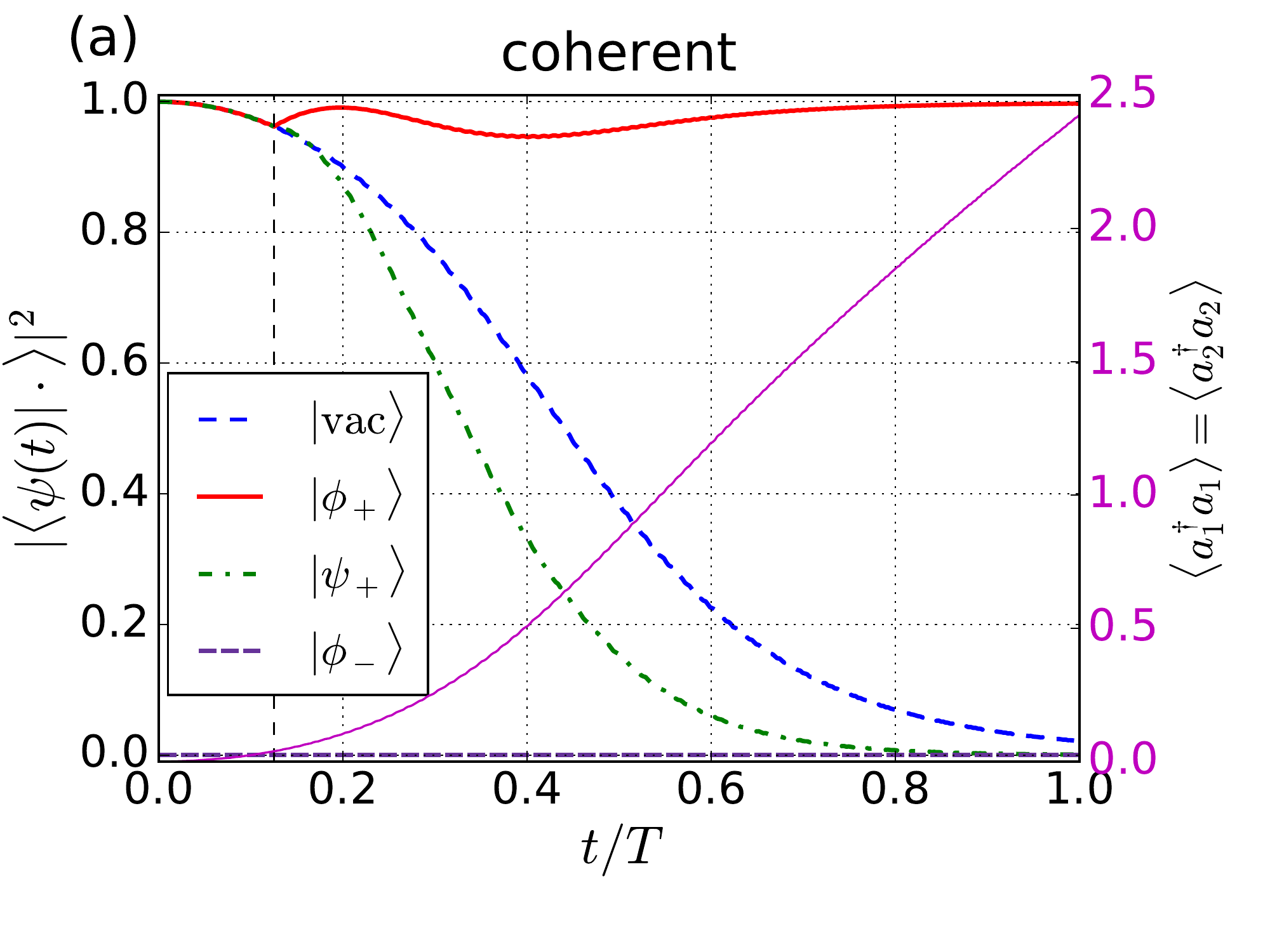}\hfill\includegraphics[width=\columnwidth]{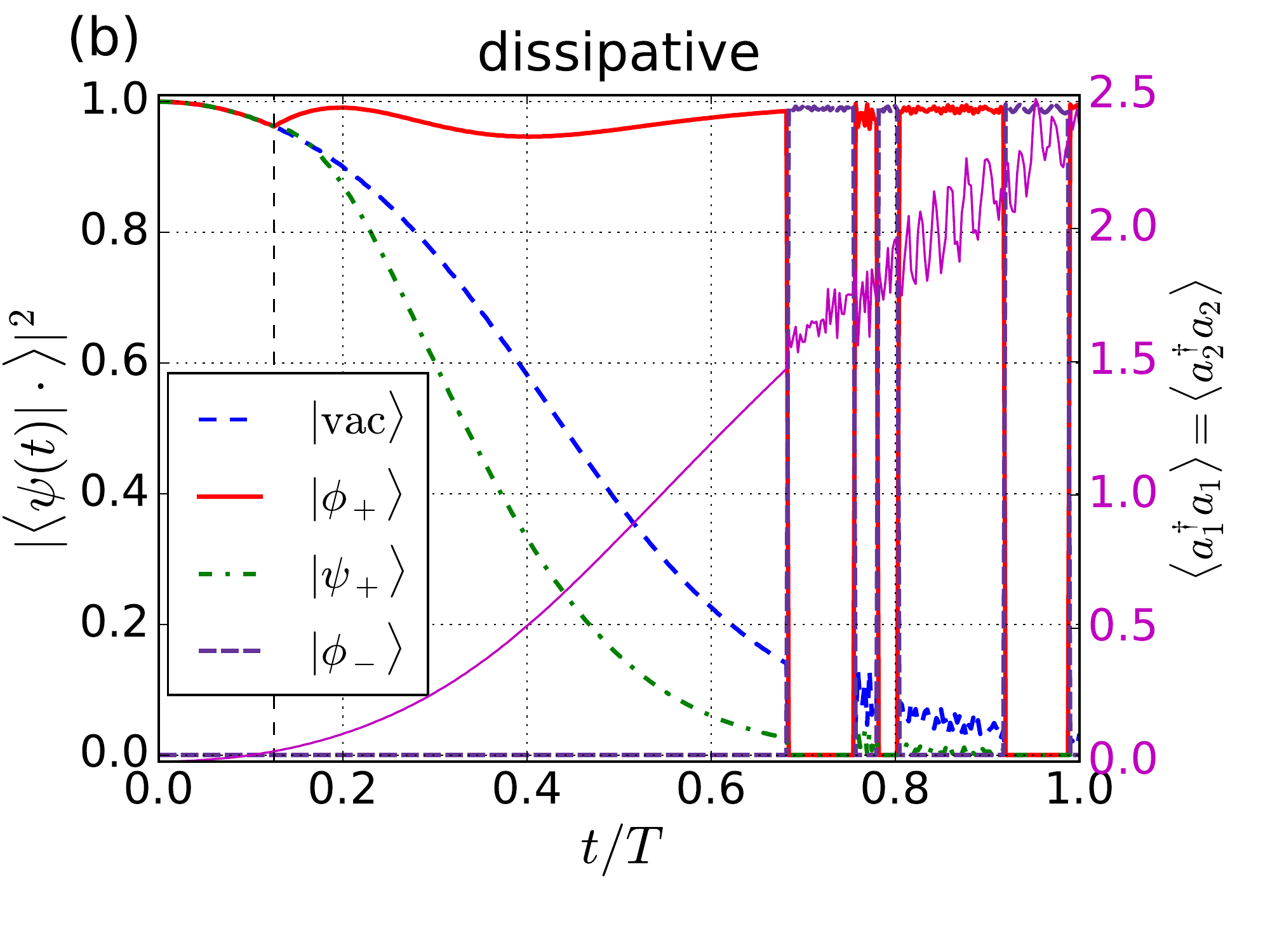}\caption{Comparison
  between coherent and dissipative quantum annealing for two anti-ferromagnetically coupled
  equal KPOs. The state
  $\ket{\psi(t)}$ is obtained by numerically solving the
  Schroedinger equation with Hamiltonian~(\ref{eq:4}). Shown are the
  fidelities with respect to the vacuum $\ket{\rm vac}=\ket{0,0}$ as
  well as the three states
  $\ket{\phi_\pm}=(\ket{\alpha,-\alpha}\pm\ket{-\alpha,\alpha})/\sqrt{2}$
  and
  $\ket{\psi_{+}}=(\ket{\alpha,\alpha}+\ket{-\alpha,-\alpha})/\sqrt{2}$. The
  amplitude $\alpha$ is given by Eq.~(\ref{eq:6}). {\bf (a)} Without dissipation: $\kappa=0$. The system evolves from the vacuum $\ket{\rm
    vac}=\ket{0}\ket{0}$ (dashed blue line) at $t=0$ to the even parity cat state
  $\ket{\phi_+}$
  (full red line) at $t=T$. The latter state encodes the ground state of the corresponding
  anti-ferromagnetic Ising model ($J<0$). The population of the odd photon number parity state
  $\ket{\phi_-}$ remains zero (dashed purple line). The bifurcation dynamics is clearly visible as a kink
  of the population of $\ket{\phi_+}$, when the drive
  strength reaches the threshold value $\varepsilon_{\rm soft}=\frac{1}{2}\sqrt{(\Delta+|J|)^2+\left( \frac{\kappa}{2} \right)^2}$ (vertical thin dashed
  black line). {\bf (b)} With dissipation:
  $\kappa=0.01\,{\rm MHz}$. Shown is a quantum trajectory with six jumps
  obtained from a Monte Carlo simulation of the dissipative
  dynamcics. A photon loss event induces a transition between the even and odd
  photon number parity cat states. Note however, that both $\ket{\phi_+}$ and
  $\ket{\phi_-}$ encode correctly the anti-ferromagnetic Ising spin
  correlations. Also note the absence of jumps
  below threshold, where the average photon number (thin magenta line)
  is close to zero. The
  parameter values used in both simulations are:
  $\Delta=-1\,{\rm MHz}$,
  $J=-0.5\,{\rm MHz}$, $K=0.7\,{\rm MHz}$, $T=400\,{\rm \mu s}$,
  $\varepsilon_{\rm MAX}=2.0\,{\rm MHz}$ and
  $\varepsilon(t)=\varepsilon_{\rm MAX}(t/T)$.\label{fig:qim_adiabatic}}
\end{figure*}
 frustration that is intimately related to spin
glass physics~\cite{Santoro-2002, Venturelli-2015}. Anti-ferromagnetic
coupling can be achieved by substituting the
shunt inductor with a large area Josephson junction and by biasing the
loop created by the $N$ KPOs and the shunt with half a flux
quantum (see Methods). Under the condition that $NE_J^{(0)}< E_J^{(n)}$, where
$E_J^{(0)}$ denotes the Josephson energy of the shunt junction, the
latter effectively acts as a {\em negative} inductor and the coupling
matrix elements become~\cite{supp_mat} $J_{nm}\simeq
-(E_J^{(0)}/2)\sqrt{Z_nZ_m}$. We remark that a
$\pi$-junction~\cite{Ryazanov-2001, Gingrich-2016, Kontos-2002} shunt, provides an alternative to realize anti-ferromagnetic couplings. Tunability of the matrix elements is
enabled by connecting a tunable capacitor in parallel with the split
junction of each KPO. While tunable
high-Q capacitors in the microwave regime are not yet part of the standard toolbox of cQED, their
development is an active area of research~\cite{Park-2004,
  Cicak-2010}. Even without this tunability, a proof-of principle
experiment is within reach of current technology.

\section*{Robustness to dissipation}
The dominant decoherence source in this system is energy dissipation
that occurs due to internal losses in the device or via the capacitively coupled
readout lines (see Fig.~\ref{fig:setup} and~\cite{supp_mat}). We describe this dissipation
by including photon losses with rate $\kappa$ using a standard Lindblad master equation
\begin{align}\label{eq:3}
\dot{\bm\rho}=-i\left( \bm H_{\rm NH}\bm\rho -\bm\rho\bm H_{\rm NH} \right)+\kappa\sum_{n=1}^N \bm
  a_n^{}\bm\rho\bm a_n^{\dagger}.
\end{align}
Here $\bm H_{\rm NH}=\bm H_N-i\frac{\kappa}{2}\sum_{n=1}^N\bm
  a_n^{\dagger}\bm a_n^{}$ and
$\bm H_N$ is given by Eq.~(\ref{eq:4}). This way of writing
the master equation emphasizes the two different aspects of photon
loss: The first term on the right hand side represents a non-unitary
but deterministic evolution of the state, while the second term on the
right hand side represents the stochastic, i.e. non-deterministic
``jump'' action of an annihilation operator on the state~\cite{Wiseman-2010}.

A key property of the continuous variable Ising encoding is its
robustness to photon loss. This robustness can be illustrated with the
simple case of two coupled identical KPOs (see
Fig.~\ref{fig:annealing}). The disspation modifies the
thresholds of the soft and hard modes~\cite{supp_mat} as $\varepsilon_{\rm
  soft/hard}=\frac{1}{2}\sqrt{(\Delta\pm |J|)^2+\left( \frac{\kappa}{2}
  \right)^2}$. If the drive strength is varied adiabatically,
then under the action of $\bm H_{\rm NH}$, the wavefunction of the
system splits deterministically into an equal
superposition between the two maxima of the potential to remain in the
instantaneous steady state~\cite{Venuti-2016}. Note that this
evolution preserves the purity of the state. In the original basis then, the two-mode vacuum evolves into a two-mode cat state with even photon
number parity (see
Fig.~\ref{fig:qim_adiabatic}~(a)):
\begin{align*}
\ket{0,0}\rightarrow\begin{cases}(\ket{\alpha,\alpha}+\ket{-\alpha,-\alpha})/\sqrt{2},&
  {\rm for}\quad J>0,\\
(\ket{\alpha,-\alpha}+\ket{-\alpha,\alpha})/\sqrt{2},& {\rm for}\quad J<0\end{cases}.
\end{align*}
Far above threshold, the amplitude can be found from the solution of the
semi-classical equations of motion~(see Methods)
\begin{align}\label{eq:6}
\alpha=e^{i\phi}\sqrt{\frac{\sqrt{4\varepsilon^2-\left(
                    \frac{\kappa}{2} \right)^2}+\Delta+|J|}{2K}},
\end{align}
with
$\phi=-\frac{1}{2}\arctan\left[\frac{\kappa}{\sqrt{16\varepsilon^2-\kappa^2}}\right]$.
\begin{figure*}[ht]
\includegraphics[width=\textwidth]{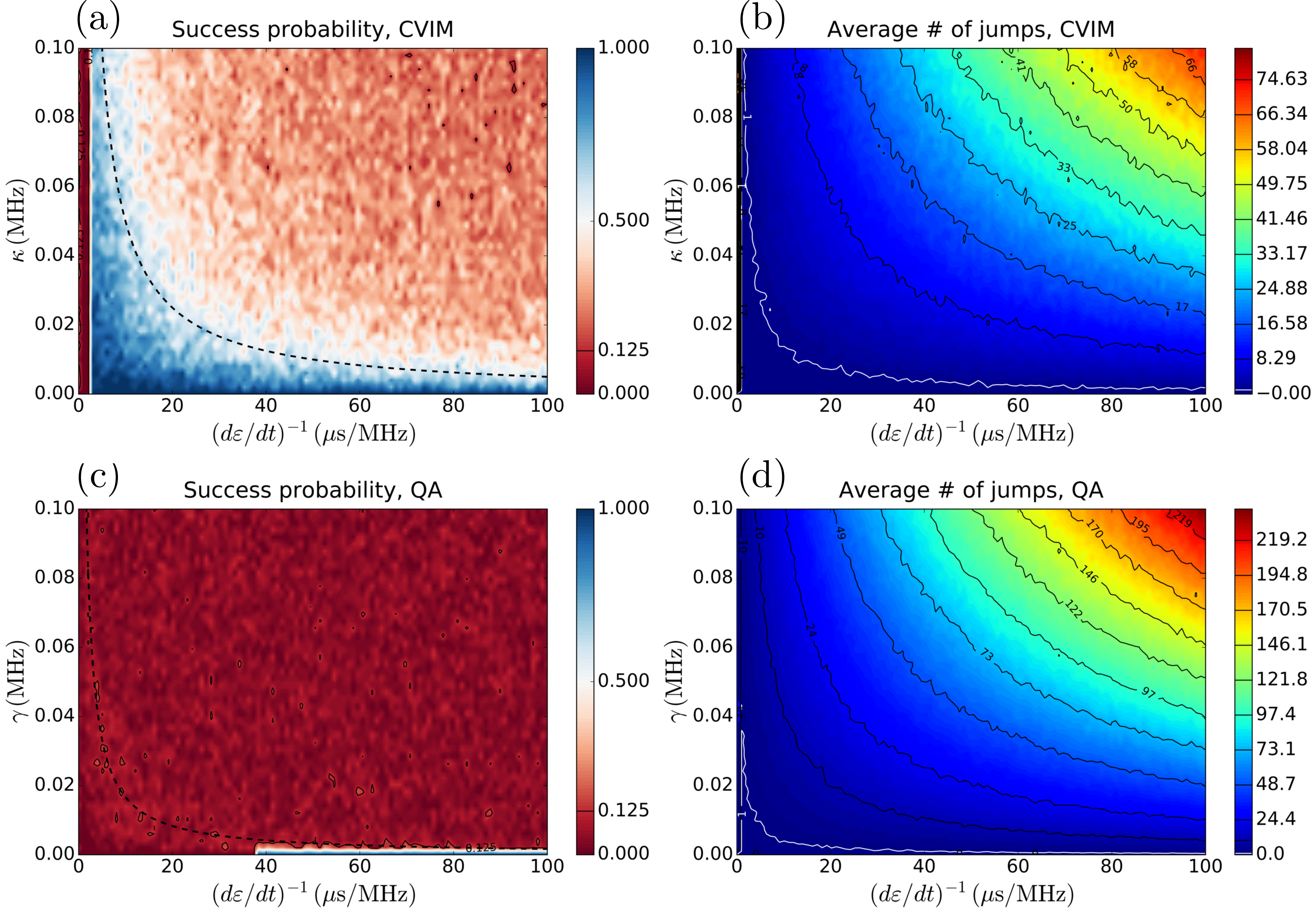}~\caption{Comparison
  of performance between the continuous variable Ising machine (CVIM) ({\bf (a)} and {\bf (b)}) and a standard discrete
  qubit based quantum annealer (QA) ({\bf (c)} and {\bf (d)}). Panels
  {\bf (a)} and {\bf (c)} show the success probability of
  the number partitioning problem with set $S=\{4,5,6,7\}$ as a
  function of the ramp rate $d\varepsilon/dt$ and the photon loss
  rate $\kappa$ in {\bf (a)} or dephasing rate $\gamma$ in {\bf
    (c)}. Panels {\bf (b)} and {\bf (d)} show the corresponding mean
  number of jump events (photon loss for {\bf (b)} and dephasing
  events for {\bf (d)}). While the success probability of the QA drops sharply
  already after a single (on average) dephasing event (see region
  within white contour in panel {\bf (d)}), the CVIM still
  succeeds with probabilities $>0.5$ even when more than one photon
  has been lost (see region within the white contour in panel {\bf
    (b)}). Also, the adiabatic ramp rate threshold for the CVIM
  is substantially higher than for the QA. Finally, the
  success probability for the CVIM is typically above that for a random guess
  $\sim 1/8=0.125$,
  in the entire region shown above the adiabatic threshold, while the
  success probability of the QA quickly drops below the random guess
  value delimited by thin solid black contour lines in {\bf (a)} and
  {\bf (c)}. The parameter values for the CVIM
  simulations {\bf (a)} and {\bf (b)} are: $\Delta=-1.5\,{\rm MHz}$,
  $K=0.6\,{\rm MHz}$, $\varepsilon_{\rm
    MAX}=2\,{\rm MHz}$ and $\varepsilon(t)=\varepsilon_{\rm MAX}\left(
    t/T \right)$ for $T$ in the range $(0, 200)\,{\rm \mu s}$. The parameters for
  the QA simulations {\bf (c)} and {\bf (d)} are:
  $\varepsilon(t)=\varepsilon_{\rm MAX}(1-t/T)$ with $\varepsilon_{\rm
  MAX}=6\,{\rm MHz}$ and $T$ in the range $(0, 600)\,{\rm \mu s}$. The dashed
black curves in {\bf (a)} and {\bf (c)} indicate the points where
$\kappa T = 1$ respectively $\gamma T=1$. Each point in all figures
corresponds to an average over $40$ trajectories.\label{fig:NPP}}
\end{figure*}

Let us now turn to the jump part of the dissipative process (see Eq.~(\ref{eq:3})).
Below threshold, the average photon number $\bar n$ is low
(see Fig.~\ref{fig:qim_adiabatic}). Therefore, the probability of a
photon loss event below threshold is strongly
suppressed and consequently the impact of dissipation on the annealing
process is reduced. Above threshold, the jumps lead to
the random switching of the photon number parity
(Fig.~\ref{fig:qim_adiabatic} (b)). In the ensemble averaged picture this results in
dephasing of the pure states into mixtures of two-mode coherent states i.e., $(\ket{\alpha,\alpha}+\ket{-\alpha,-\alpha})/\sqrt{2}\xrightarrow{\bm
a, \bm b}{(\ket{\alpha,\alpha}\bra{\alpha,\alpha}+\ket{-\alpha,-\alpha}\bra{-\alpha,-\alpha})/2}$, as
well as $(\ket{\alpha,-\alpha}+\ket{-\alpha,\alpha})/\sqrt{2}\xrightarrow{\bm
a, \bm b}{(\ket{\alpha,-\alpha}\bra{\alpha,-\alpha}+\ket{-\alpha,\alpha}\bra{-\alpha,\alpha})/2}$.
Crucially though, photon loss above threshold, does not
corrupt the Ising spin correlations, i.e. for $J>0$ ($J<0$), the
two oscillators still oscillate in-phase (with opposite phase)
(Figs.~\ref{fig:annealing}~(d) and~\ref{fig:qim_adiabatic}). This robustness of the annealing process to dissipation, is to
be contrasted with a conventional discrete qubit based implementation
of a quantum annealer, where initially the state of the annealer is a fragile
coherent superposition of all possible spin configurations and the
impact of qubit dephasing in the Ising basis before the
avoided crossing strongly reduces the population of the ground
state~\cite{Boixo-2016,supp_mat}.

\section*{Application to number partitioning}

To demonstrate the capabilities of our proposed device, we have simulated an instance of the
NPP with $N=4$ oscillators using a Monte Carlo quantum trajectory
algorithm~\cite{Qutip-2012a}. The NPP is defined by the set $A=\{4,5,6,7\}$, for which a fair partition is $F=\{4,7\}$ and
$\bar F=\{5,6\}$. 
To encode the corresponding Ising problem
into our circuit we set the coupling matrix elements as
$J_{ij}=-J_0A_iA_j$ and $J_{ii}=0$.
Here $J_0>0$ is a scale factor that leaves the Ising
problem invariant and which is convenient to satisfy
the physical constraints on the coupling strengths. We choose
$J_0=1/{\rm max}|J_{ij}|=1/42$. The two Ising spin
configurations that satisfy this NPP are $\{s_1=s_4=1, s_2=s_3=-1\}$, and
the configuration with all spins flipped. Fig.~\ref{fig:NPP}~(a) shows the
success probability as a function of $(d\varepsilon/dt)^{-1}=T/\varepsilon_{\rm MAX}$ and the
photon loss rate $\kappa$. Fig.~\ref{fig:NPP} (b) shows the
corresponding average number of jump events. Each point in both
figures represents an average over
$40$ trajectories. A particular run of the optimizer is deemed
successful if the phase correlations between each pair of
oscillators, as measured by ${\rm sign}\left\{ \cos\left( {\rm arg}[ \braket{\bm a_i^{\dagger}\bm
  a_j^{}} ]\right) \right\}$, is equal to $s_is_j$, at the final drive strength value
$\varepsilon_{\rm MAX}$. For comparison, Fig.~\ref{fig:NPP} (c) shows the success
probability, obtained for the same NPP, when simulating a standard quantum annealer
consisting of four qubits that are subject to dephasing with rate
$\gamma$ in the
basis that diagonalizes the Ising terms. Clearly, the optimization
in the CVIM is more robust. Success
probabilities above $0.5$ are obtained even when the average number of
photons lost during the annealing is larger than one
(Fig.~\ref{fig:NPP} (b)). In
contrast, the optimization with the spin-based quantum annealer fails
already after a single dephasing event (Fig.~\ref{fig:NPP} (d)). This can be understood by
observing that in the initial state of the annealer, where the spins
are polarized along $\bm\sigma^x_i$ for $i\in\{1,2,3,4\}$, a dephasing error described by
$\bm\sigma^z_i$ induces transitions out of the ground state to
excited states that are not adiabatically connected with the solution
of the optimization problem~\cite{supp_mat}.

In conclusion, we have proposed and investigated both analytically and
numerically the implementation of a quantum
Ising optimization machine with superconducting circuits, which solves
two of the major challenges in the field. First, we show that flux
quantization enables the realization of all-to-all connectivity among
the Ising spins,  without overhead, i.e., $N$ oscillators are sufficient to encode $N$ Ising
spins with full connectivity. Second, we show that with this continuous variable encoding of the Ising minimization problem,
quantum annealing succeeds with high probability in a dissipative
regime with high error rate where
conventional discrete qubit based quantum annealing breaks down. Our
results open up new perspectives for quantum optimization.

\paragraph{Acknowledgments} The numerical calculations were performed in
a parallel computing environment at sciCORE (\url{http://scicore.unibas.ch/}) scientific computing core facility
at University of Basel. S.~E.~N. acknowledges financial support from the
Swiss NSF. R.~P.~T. was financially supported
by the Swiss SNF and the NCCR Quantum Science and Technology.
\paragraph{Author contributions} 
S.~E.~N. developed the concepts, carried out the calculations, wrote the numerical code and the
manuscript and contributed to the interpretation of the results. R. P. T. contributed to the 
calculations, the interpretation of the results and the manuscript and N. L. contributed to the 
numerics, the interpretation of the results and the manuscript.

\section*{Methods}
\subsection*{Two coupled KPOs}
The Hamiltonian of two coupled KPOs in the basis of the
symmetric/anti-symmetric modes $\bm d$ and $\bm c$ reads
\begin{align*}
K(\bm a^{\dagger}\bm a^{\dagger}\bm a\bm a+\bm b^{\dagger}\bm
  b^{\dagger}\bm b\bm b)&=\frac{K}{2}(\bm d^{\dagger}\bm d^{\dagger}\bm
  d\bm d +\bm c^{\dagger}\bm c^{\dagger}\bm c\bm c)+K\bm d^{\dagger}\bm
  d\bm c^{\dagger}\bm c\\
&+\frac{K}{2}(\bm d^{\dagger}\bm d^{\dagger}\bm c\bm c+\bm
  c^{\dagger}\bm c^{\dagger}\bm d\bm d).
\end{align*}
Given that the frequency of the symmetric/anti-symmetric modes are
$\Delta\pm J$, the terms on the second line rotate with frequency $2J$ and if $K\ll
4|J|$ they can be neglected in rotating wave approximation.
\subsection*{Semi-classical equations of motion}
The semi-classical equations of motion are obtained by replacing the
quantum operators $\bm a$ and $\bm b$ by complex functions in the
Heisenberg-Langevin equations of motion. They read
\begin{align*}
i\dot\alpha &= (\Delta -2K|\alpha|^2)\alpha+J\beta
  +2\varepsilon\alpha^*-i\frac{\kappa}{2}\alpha,\\
i\dot\beta &= (\Delta -2K|\beta|^2)\beta+J\alpha
  +2\varepsilon\beta^*-i\frac{\kappa}{2}\beta.
\end{align*}
A linear stability analysis of the solutions of these equations is
provided in Ref.~\cite{supp_mat}.

\subsection*{Anti-ferromagnetic coupling}
The potential energy of the system with a large area Josephson
junction shunt and a half-flux quantum flux bias $\Phi_e=\Phi_0/2$ is
\begin{align*}
\mathcal{U}=E_J^{(0)}\cos\left( \sum_{n=1}^N\varphi_n \right)-\sum_{n=1}^NE_J^{(n)}\cos\left( \varphi_n \right).
\end{align*}
For small oscillations, the potential minimum satisfies the transcendental equation~\cite{supp_mat}
\begin{align*}
\varphi_n\simeq \frac{E_J^{(0)}}{E_J^{(n)}}\sin(N\varphi_n).
\end{align*}
When $NE_J^{(0)}< E_J^{(n)}$, the only solution of this equation is
$\varphi_n=0$. Expanding the shunt potential around this classical
minimum and quantizing yields the anti-ferromagnetic interaction term
given in the main text~\cite{supp_mat}.

\bibliography{biblio_short}
\bibliographystyle{apsrev}

\cleardoublepage
\setboolean{@twoside}{false}


\section*{Supplementary material}

# Supplementary Material for
# "Robust quantum optimizer with full connectivity"

Simon E. Nigg, Niels Lörch and Rakesh P. Tiwari
(Dated: September 20, 2016)

**CONTENTS**

## I. DERIVATION OF THE CIRCUIT HAMILTONIAN

We start by describing an individual split Josephson junction with ac flux bias. We then discuss the condition for achieving anti-ferromagnetic all-to-all coupling between $N$ Kerr parametric oscillators (KPOs) connected in series and derive the Hamiltonian of the quantum Ising machine described in the main text.

### A. AC modulated split Josephson junction as a parametric oscillator

Consider a small loop of superconductor interrupted twice by a normal tunnel junction forming two identical Josephson junctions. The ring is assumed to be small enough so that its inductance is dominated by the Josephson inductances. The Lagrangian is then

$$\mathcal{L}_0 = \frac{1}{2}C_J(\dot{\varphi}_1^2 + \dot{\varphi}_2^2) + E_J\left(\cos\varphi_1 + \cos\varphi_2\right). \tag{1}$$

Here, for convenience, we have chosen units such that the flux quantum $\Phi_0 = h/(2e) = 2\pi$. The parameters $C_J$ and $E_J$ denote the capacitance and Josephson energy of each Josephson junction. The variables $\varphi_1$ and $\varphi_2$ are not independent.

If a flux $\Phi$ is threaded though the loop, then flux quantization requires that $\varphi_1 + \varphi_2 = \Phi$. It is convenient to change variables as

$$\chi_i = \varphi_i - \frac{\Phi}{2}. \tag{2}$$

Then $\chi_1 + \chi_2 = 0$ and $\chi_1 - \chi_2 = \varphi_1 - \varphi_2 \equiv 2\theta$ and the Lagrangian becomes

$$\mathcal{L}_0 = \frac{1}{2}C\dot{\theta}^2 + 2E_J \cos(\Phi/2) \cos\theta, \tag{3}$$

with $C = 2C_J$. Here we have used that $d(\chi_1 + \chi_2)/dt = 0$ and we have neglected a time dependent term $\frac{1}{4}C_J\dot{\Phi}^2$, which is independent of the coordinates and does not affect the Euler Lagrange equations. Note that this lumped element circuit describes a parametric oscillator and is essentially equivalent to the one used to describe the Josephson parametric amplifier[1,2].

A tunable capacitor, required to achieve tunable couplings (see Methods in the main text), can be added in parallel with the Josephson capacitance as illustrated in Fig. 1. The total capacitance is then the sum of the tunable capacitance and the Josephson capacitances.

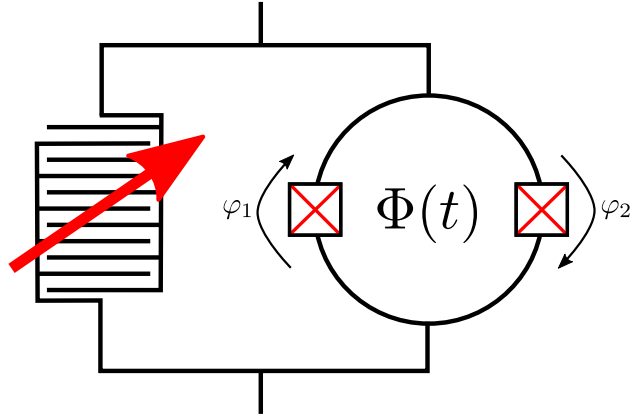

FIG. 1. Illustration of a SQUID based Kerr parametric oscillator circuit. The square boxes with red crosses represent the Josephson junctions while the interdigitated diagram on the left represents a tunable capacitor.

The Hamiltonian function is obtained as usual by introducing the momentum conjugate to the variable $\theta$ as

$$q \equiv \frac{1}{2e} \frac{\partial \mathcal{L}_0}{\partial \dot{\theta}} = \frac{C}{2e}\dot{\theta}. \tag{4}$$

Then a Legendre transformation yields

$$\mathcal{H}_0 = 2eq\dot{\theta} - \mathcal{L}_0 = 4E_C q^2 - 2E_J \cos(\Phi/2) \cos\theta, \tag{5}$$

where we have defined the single electron charging energy $E_C = e^2/(2C)$.

### 1. Ac flux modulation

Next, we consider a monochromatic flux modulation of the form

$$\Phi(t)/2 = \Phi^{\text{dc}} + \delta\Phi(t), \tag{6}$$

with $\delta\Phi(t) = \delta\Phi^{\text{ac}} \sin(\Omega t + \phi)$. Then we have

$$\cos(\Phi(t)/2) = \cos(\Phi^{\text{dc}}) \cos\left(\delta\Phi^{\text{ac}} \sin(\Omega t + \phi)\right) - \sin(\Phi^{\text{dc}}) \sin\left(\delta\Phi^{\text{ac}} \sin(\Omega t + \phi)\right) \tag{7}$$

$$= \cos(\Phi^{\text{dc}}) \left\{ J_0\left(\delta\Phi^{\text{ac}}\right) + 2 \sum_{m=1}^{\infty} J_{2m}\left(\delta\Phi^{\text{ac}}\right) \cos\left[2m\left(\Omega t + \phi\right)\right] \right\} \tag{8}$$

$$- 2 \sin(\Phi^{\text{dc}}) \sum_{m=0}^{\infty} J_{2m+1}\left(\delta\Phi^{\text{ac}}\right) \sin\left[(2m+1)\left(\Omega t + \phi\right)\right]. \tag{9}$$

If $|\delta\Phi^{ac}| \ll 1$, then we can keep in the sums only terms with $J_0, J_1$, yielding

$$\cos(\Phi(t)/2) \simeq \cos(\Phi^{dc})J_0(\delta\Phi^{ac}) - 2\sin(\Phi^{dc})J_1(\delta\Phi^{ac})\sin(\Omega t + \phi). \tag{10}$$

It is convenient to separate the Hamiltonian into an explicitly time independent and an explicitly time dependent part as $\mathcal{H}_0 = \mathcal{H}_1 + \mathcal{H}_2(t)$, where

$$\mathcal{H}_1 = 4E_C q^2 - 2E_J \cos(\Phi^{dc})J_0(\delta\Phi^{ac})\cos\theta, \tag{11}$$

and

$$\mathcal{H}_2(t) = 4E_J \sin(\Phi^{dc})J_1(\delta\Phi^{ac})\sin(\Omega t + \phi)\cos\theta. \tag{12}$$

### 2. Expansion of the Josephson nonlinearity

We are ultimately interested in the transmon regime[3] where $E_J \gg E_C$. In this case we can expand the $\cos\theta$ term around the local minimum of the classical potential $\theta = 0$, and keep only the leading nonlinearity. Furthermore, since we focus on the case of weak ac modulation, where $\delta\Phi^{ac} \ll \Phi_0$ (in physical units), we can neglect the nonlinear part of the time dependent term and approximate $J_0(x) \simeq 1$ and $J_1(x) \simeq x/2$. Therefore, dropping c-number terms which do not contribute to the dynamics, we have

$$\mathcal{H}_1 \simeq 4E_C q^2 + E_J \cos(\Phi^{dc})\theta^2 - \frac{E_J}{12}\cos(\Phi^{dc})\theta^4, \tag{13}$$

$$\mathcal{H}_2 \simeq -E_J \sin(\Omega t + \phi)\sin(\Phi^{dc})\delta\Phi^{ac}\theta^2. \tag{14}$$

### 3. Canonical quantization

The quantum model is introduced by replacing the generalized coordinates $\theta$ and $q$ by canonically conjugate operators $\boldsymbol{\theta}$ and $\boldsymbol{q}$ with commutator[3] $[\boldsymbol{\theta}, \boldsymbol{q}] = i$. It is then convenient to work in the basis which diagonalizes the quadratic part of $\boldsymbol{H}_1$. This is achieved by introducing the bosonic creation and annihilation operators $\boldsymbol{a}^\dagger$ and $\boldsymbol{a}$, with commutator $[\boldsymbol{a}, \boldsymbol{a}^\dagger] = 1$, defined as

$$\boldsymbol{\theta} = \sqrt{\frac{Z}{2}}\left(\boldsymbol{a} + \boldsymbol{a}^\dagger\right), \tag{15}$$

$$\boldsymbol{q} = -i\sqrt{\frac{1}{2Z}}\left(\boldsymbol{a} - \boldsymbol{a}^\dagger\right). \tag{16}$$

Here we have defined the mode impedance

$$Z = \sqrt{\frac{4E_C}{E_J \cos(\Phi^{dc})}}. \tag{17}$$

Thus the Hamiltonian becomes

$$\boldsymbol{H}_1 = \omega \boldsymbol{a}^\dagger \boldsymbol{a} - \frac{E_C}{12}\left(\boldsymbol{a} + \boldsymbol{a}^\dagger\right)^4, \tag{18}$$

$$\boldsymbol{H}_2 = -\frac{E_J}{2}Z \sin(\Omega t + \phi)\sin(\Phi^{dc})\delta\Phi^{ac}\left(\boldsymbol{a} + \boldsymbol{a}^\dagger\right)^2. \tag{19}$$

Here we have defined the oscillator harmonic frequency

$$\omega = 4\sqrt{E_C E_J \cos(\Phi^{dc})}. \tag{20}$$

Next, we use the rotation wave approximation (RWA) twice. The first time, to eliminate the counter-rotating terms in the quartic terms in $\boldsymbol{H}_1$. Normal ordering yields

$$:(\boldsymbol{a} + \boldsymbol{a}^\dagger)^4: = \boldsymbol{a}^{4\dagger} + 6\boldsymbol{a}^{2\dagger} + 4\left(\boldsymbol{a}^{3\dagger}\boldsymbol{a}\right) + 6\left(\boldsymbol{a}^{2\dagger}\boldsymbol{a}^2\right) + 4\left(\boldsymbol{a}^\dagger \boldsymbol{a}^3\right) + 12\left(\boldsymbol{a}^\dagger \boldsymbol{a}\right) + \boldsymbol{a}^4 + 6\boldsymbol{a}^2 + 3 \tag{21}$$

Hence the non-rotating terms are $6a^{\dagger 2}a^2$ and $12a^\dagger a$. The latter leads to a small renormalization of the oscillator frequency by an amount $\Delta\omega = -E_C$. Henceforth, we shall absorb this term in an implicit redefinition of the oscillator frequency. The second application of the RWA is with respect to the drive. Here we specifically consider a drive frequency close to twice the oscillator frequency $\Omega \simeq 2\omega$. Then if we furthermore choose the drive phase $\phi = -\pi/2$, the only resonant terms in $H_2$ are the ones proportional to $a^2 e^{i\Omega t}$ and $a^{\dagger 2} e^{-i\Omega t}$. Putting it all together, we obtain the Hamiltonian of the KPO

$$H = \omega a^\dagger a - \frac{E_C}{2} a^\dagger a^\dagger a a + \frac{E_J}{4} Z \sin(\Phi^{\text{dc}}) \delta\Phi^{\text{ac}} \left( a^2 e^{i\Omega t} + a^{\dagger 2} e^{-i\Omega t} \right). \tag{22}$$

### B. Shunted chain of parametric oscillators

Next, we consider a chain of KPOs shunted by a large area Josephson junction and revert to the notation used in the main text denoting with $\varphi_n$ the phase drop across oscillator $n \in \{1, \ldots, N\}$. The phase drop across the shunt junction is denoted with $\varphi_0$. The total Lagrangian of the shunted chain of KPOs depicted in Fig. 1 of the main text is

$$\mathcal{L} = E_J^{(0)} \cos(\varphi_0) + \sum_{n=1}^{N} \mathcal{L}_n, \tag{23}$$

where $E_J^{(0)}$ denotes the Josephson energy of the shunt and $\mathcal{L}_n$ are the Lagrangians of the individual KPOs (3) with $\theta \to \varphi_n$. Here we have neglected the capacitive energy of the large area shunt junction, assuming that its capacitance is sufficiently large such that $E_C^{(0)}/E_J^{(0)} \approx 0$. The phase variables $\varphi_0, \varphi_1, \ldots, \varphi_N$ are not independent. Flux quantization requires that

$$\varphi_0 = \Phi_e - \sum_{n=1}^{N} \varphi_n. \tag{24}$$

Here $\Phi_e$ denotes the flux bias through the loop formed by the chain and the shunt.

### C. Minima of the potential energy with half flux quantum flux bias

When deriving the quantum Hamiltonian of the circuit, it is important to expand around a classical minimum of the potential. In terms of the $N$ independent variables $\varphi_1, \ldots, \varphi_N$, the potential energy is

$$\mathcal{U} = -E_J^{(0)} \cos\left(\Phi_e - \sum_{n=1}^{N} \varphi_n\right) - 2 E_J \sum_{n=1}^{N} \cos\left(\Phi_n^{\text{dc}}\right) J_0(\delta\Phi_n^{\text{ac}}) \cos(\varphi_n). \tag{25}$$

The location of the minima depends on the flux bias $\Phi_e$. When $\Phi_e = 0$, then the point $\varphi_1 = \varphi_2 = \cdots = \varphi_N = 0$ is a local minimum. Of special interest for implementing anti-ferromagnetic coupling is the case $\Phi_e = \pi$, corresponding to half a flux quantum. In this case the potential energy is

$$\mathcal{U} = E_J^{(0)} \cos\left(\sum_{n=1}^{N} \varphi_n\right) - \sum_{n=1}^{N} E_J^{(n)} \cos(\varphi_n), \tag{26}$$

where we have defined $E_J^{(n)} = 2 E_J \cos(\Phi_n^{\text{dc}}) J_0(\delta\Phi_n^{\text{ac}})$. The minima of the potential satisfy

$$\partial_{\varphi_n} \mathcal{U} = 0 \Leftrightarrow E_J^{(n)} \sin(\varphi_n) = E_J^{(0)} \sin\left(\sum_{n=1}^{N} \varphi_n\right). \tag{27}$$

This implies that $\varphi_n - \varphi_m = 2\pi k$ or $\varphi_n + \varphi_m = (2k+1)\pi$ with $k \in \mathbb{Z}$. Let us focus on the case where $|\varphi_n| \ll \pi$, which is shown to be consistent a posteriori. In this case we can only have the branch of solutions satisfying $\varphi_n - \varphi_m = 2\pi k$. Then

$$\varphi_n \simeq \sin(\varphi_n) = \frac{E_J^{(0)}}{E_J^{(n)}} \sin(N\varphi_n). \tag{28}$$

If now we further require that $NE_J^{(0)} < E_J^{(n)}$, then only one solution of the transcendental equation (28) exists namely $\varphi_n = 0$. Physically, this condition means that the potential of the shunt junction is much softer than the potential of the chain junctions. Therefore, in the ground state it is preferable to minimize the energy stored in the stiff chain junctions while allowing a larger deviation of $\varphi_0$ across the soft shunt junction away from its equilibrium value $\pi$.

### D. Expansion of the Josephson potential of the shunt: All-to-all coupling

Having established the condition for a local minimum of the potential at $\varphi_1 = \varphi_2 = \cdots = \varphi_N = 0$ when $\Phi_e = \pi$, we next expand the potential energy around this point up to the leading non-linearity. This procedure is very similar to the one discussed in I A 2 and we only point out the differences here. With the flux bias $\Phi_e = \pi$, the shunt contributes a term to the Hamiltonian

$$\mathcal{H}_S = E_J^{(0)} \cos\left(\sum_{n=1}^{N} \varphi_n\right). \tag{29}$$

Since $NE_J^{(0)} < E_J$, we neglect the nonlinear contribution and expanding the cosine to leading order. Suppressing an unimportant c-number term we then have

$$\mathcal{H}_S \simeq -\frac{E_J^{(0)}}{2} \sum_{n=1}^{N} \varphi_n^2 - \frac{E_J^{(0)}}{2} \sum_{n \neq m} \varphi_n \varphi_m. \tag{30}$$

The full Hamiltonian is thus (see Eqs. (13) and (14))

$$\mathcal{H} = 4 \sum_{n=1}^{N} E_C^{(n)} q_n^2 - \sum_{n=1}^{N} \left[E_J^{(n)} \cos(\Phi_n^{\rm dc}) - \frac{E_J^{(0)}}{2}\right] \varphi_n^2 - \sum_{n=1}^{N} \frac{E_J^{(n)}}{12} \cos(\Phi_n^{\rm dc}) \varphi_n^4 - \sum_{n=1}^{N} \sin(\Omega_n t + \phi_n) \sin(\Phi_n^{\rm dc}) \delta\Phi_n^{\rm ac} \varphi_n^2 - \frac{E_J^{(0)}}{2} \sum_{n \neq m} \varphi_n \varphi_m. \tag{31}$$

#### 1. Canonical quantization

Quantization is similar to the case of single parametric oscillator in Section. I A 3. The only difference is that the mode impedances and frequencies also depend on $E_J^{(0)}$ as follows

$$\omega_n = 4\sqrt{E_C^{(n)}\left[E_J^{(n)} \cos(\Phi_n^{\rm dc}) - \frac{E_J^{(0)}}{2}\right]}, \tag{32}$$

$$Z_n = \sqrt{\frac{4E_C^{(n)}}{E_J^{(n)} \cos(\Phi_n^{\rm dc}) - \frac{E_J^{(0)}}{2}}} = \frac{8E_C^{(n)}}{\omega_n}. \tag{33}$$

Finally, choosing the drive frequencies $\Omega_n \approx 2\omega_n$, setting $\phi_n = -\pi/2$ and applying the RWA, the Hamiltonian of $N$ coupled KPOs with all-to-all anti-ferromagnetic coupling is

$$H = \sum_{n=1}^{N} \omega_n a_n^\dagger a_n + \sum_{n=1}^{N} \varepsilon_n \left(e^{i\Omega_n t} a_n^2 + e^{-i\Omega_n t} a_n^{\dagger 2}\right) + \sum_{n \neq m} J_{nm} a_n^\dagger a_m - \sum_{n=1}^{N} K_n a_n^\dagger a_n^\dagger a_n a_n, \tag{34}$$

with

$$\varepsilon_n \simeq \frac{E_J^{(n)}}{4} Z_n \sin\left(\Phi_n^{\rm dc}\right) \delta\Phi_n^{\rm ac}, \tag{35}$$

$$J_{nm} \simeq -\frac{E_J^{(0)}}{2} \sqrt{Z_n Z_m}, \tag{36}$$

$$K_n \simeq \frac{E_C^{(n)}}{2}. \tag{37}$$

## II. READOUT AND DISSIPATION

To read out the phase of each KPO at the end of the annealing procedure, one can capacitively couple each oscillator to a transmission line as shown in Fig. 1 of the main text. The phase shift of photons reflected back from the system can then be measureed using a standard homodyne detection scheme.

### A. Homodyne measurement via capacitively coupled transmission lines

In RWA, the full Hamiltonian of the system, including the quantized modes $b_n(\omega)$ of the semi-infinite transmission lines, is then $H_{\text{SB}} = H + H_{\text{TL}}$ with $H$ given in Eq. (34) and

$$H_{\text{TL}} = \sum_{n=1}^{N} \int_0^\infty d\omega \sqrt{\frac{\kappa_n(\omega)}{2\pi}} \left[ b_n^\dagger(\omega) a_n + a_n^\dagger b_n(\omega) \right] + \sum_n \int_0^\infty d\omega \omega b_n^\dagger(\omega) b_n(\omega). \tag{38}$$

From this model, we can derive in the usual white noise approximation $\kappa_n(\omega) = \kappa_n$, the quantum Langevin equation of motion for the oscillator modes. They read

$$\dot{a}_n(t) = -i[a_n(t), H(t)] + \frac{\kappa_n}{2} a_n(t) - \sqrt{\kappa_n} b_n^{(\text{out})}(t). \tag{39}$$

Here the output field is defined as

$$b_n^{(\text{out})}(t) = \frac{i}{\sqrt{2\pi}} \int_{-\infty}^{\infty} d\omega e^{-i\omega(t-t_1)} b_n(\omega, t_1), \tag{40}$$

with $t_1 > t$. Performing the above commutator we find

$$\dot{a}_n(t) = -i\omega_n a_n(t) - i \sum_{m \neq n} J_{nm} a_m(t) + 2iK_n a_n^\dagger a_n a_n - 2i\varepsilon_n e^{-i\Omega_n t} a_n^\dagger(t) - \frac{\kappa_n}{2} a_n(t) + \sqrt{\kappa_n} b_n^{(\text{in})}(t). \tag{41}$$

Furthermore, the standard input-output relation holds

$$b_n^{(\text{in})}(t) + b_n^{(\text{out})}(t) = \sqrt{\kappa_n} a_n(t). \tag{42}$$

Clearly an exact solution of Eqs. (41) and (42) is complicated by the presence of the Kerr nonlinearity and the long-range coupling. For any finite number of oscillators $N$, the solution below threshold can be obtained by inversion of a linear system of equations[4]. Of particular interest here is the regime far above threshold, where we can use the semi-classical approximation and replace $a_n$ with $\alpha_n = \pm|\alpha_n|e^{i\phi_n}$, the steady state amplitude of oscillator $n$. In contrast to the readout of a qubit dispersively coupled to a resonator, we do not need to drive the transmission line since the oscillators are pumped via the ac flux modulation. At zero temperature, the input is thus the vacuum noise $b_0^{(\text{in})}$. The output is then

$$b_n^{(\text{out})} = \pm \sqrt{\kappa_n} |\alpha_n| e^{i\phi_n} - b_0^{(\text{in})}. \tag{43}$$

By mixing this signal, at a beam splitter, with a strong coherent local oscillator with fixed reference phase and carrying a photon flux $\sqrt{\kappa_n} \gamma_n^{(\text{LO})}$ we have

$$b_n^{(\text{out})} = \sqrt{\kappa_n} \left( \pm r|\alpha_n| e^{i\phi_n} + t\gamma_n^{(\text{LO})} \right) - r b_0^{(\text{in})}, \tag{44}$$

where $r$ and $t$ denote the reflection and transmission amplitudes of the beam splitter. Thus we can detect the phase of the oscillator amplitude in the modulation of the output photon number

$$\int_0^T dt \langle b_n^{(\text{out})\dagger} b_n^{(\text{out})} \rangle = N_n^{(0)} \pm rt\kappa_n T \left( |\alpha_n| e^{-i\phi_n} \gamma_n^{(\text{LO})} + |\alpha_n| e^{i\phi_n} \gamma_n^{(\text{LO})*} \right). \tag{45}$$

Here $N_n^{(0)} = (|r\gamma_n^{(\text{LO})}|^2 + |t\alpha_n|^2)T \simeq |r\gamma_n^{(\text{LO})}|^2 T$, for $r \gg t$. Maximal contrast is obtained by tuning the phase of the local oscillator to $\phi_n$. Then

$$\Delta N_n^{\text{out}} = \pm rt\kappa_n T |\alpha_n \gamma_n^{(\text{LO})}|. \tag{46}$$

The coupling via the transmission lines opens up a decoherence channel via photon loss. In the main text we use a master equation (Eq. (11)) to describe such losses, which is equivalent to the quantum Langevin equation (41) with a zero temperature bath.

## III. TWO COUPLED PARAMETRIC OSCILLATORS

In the main text, we provide an intuitive picture in terms of the symmetric and anti-symmetric modes. Here we provide a complementary discussion of the dynamics of two coupled KPOs in the original basis. We consider two driven symmetric coupled KPOs. This system is described by the Hamiltonian

$$H_2 = \omega(a^\dagger a + b^\dagger b) + \varepsilon\left(a^2 e^{i\Omega t} + a^{\dagger 2} e^{-i\Omega t}\right) - K\left(a^\dagger a^\dagger aa + b^\dagger b^\dagger bb\right) + J\left(a^\dagger b + b^\dagger a\right). \tag{47}$$

Here we are interested in the case where $\Omega = 2(\omega - \Delta)$ with $|\Delta| \ll |\omega|$. Transforming to a frame rotating with the frequency $\Omega/2$, via the unitary transformation $\exp[i(\Omega/2)t(a^\dagger a + b^\dagger b)]$, we have

$$\tilde{H}_2 = -K\left(a^{\dagger 2} - \frac{\varepsilon}{K}\right)\left(a^2 - \frac{\varepsilon}{K}\right) - K\left(b^{\dagger 2} - \frac{\varepsilon}{K}\right)\left(b^2 - \frac{\varepsilon}{K}\right) + J\left(a^\dagger b + b^\dagger a\right) + \Delta\left(a^\dagger a + b^\dagger b\right) + \frac{2\varepsilon^2}{K}. \tag{48}$$

### A. Instantaneous eigenspectrum in the rotating frame

We begin by describing the low photon number part of the spectrum of (48). At zero drive strength $\varepsilon = 0$, the total number of photons shared by the two oscillators is conserved and the sector with up to one photon can be immediately diagonalized. The three states $|0,0\rangle$, $\frac{1}{\sqrt{2}}(|0,1\rangle + |1,0\rangle)$ and $\frac{1}{\sqrt{2}}(|0,1\rangle - |1,0\rangle)$, are eigenstates with corresponding eigenvalues $E_0 = 0$, $E_+ = \Delta + J$ and $E_- = \Delta - J$. Note that the corresponding energies in the "lab" frame (ignoring the zero point energy) are $E_0^{\text{lab}} = 0$, $E_\pm^{\text{lab}} = \omega \pm J$. Hence we see that if $\Delta \pm J < 0 \Leftrightarrow \Omega > 2(\omega \pm J)$, then the vacuum is guaranteed to be the state with the highest energy. The fact that the Hamiltonian (48) is not bounded from below need not alarm us if we keep in mind how it was derived by truncating the expansion of the Josephson potential and if we also keep in mind that photon losses, once included, will naturally tend to move the state towards lower photon numbers. This means however that we must now think of the instantaneous "ground state" in the rotating frame, where the Hamiltonian is time independent, as the state with highest energy instead of the more usual lowest energy state in a stationary system. For lack of a better word, we shall call this state the roof state.

The two-photon sector is spanned by the three states $|2,0\rangle$, $|1,1\rangle$ and $|0,2\rangle$. In this basis the two-photon block of the Hamiltonian at zero drive is in the rotating frame

$$H_2^{(2)} = \begin{pmatrix} 2(\Delta - K) & J & 0 \\ J & 2\Delta & J \\ 0 & J & 2(\Delta - K) \end{pmatrix}. \tag{49}$$

The three eigenstates and corresponding eigenvalues are:

| Eigenvector | Eigenvalue | Energy in "lab" frame |
|---|---|---|
| $\frac{1}{\sqrt{2}}(|2,0\rangle - |0,2\rangle)$ | $2(\Delta - K)$ | $2(\omega - K)$ |
| $\frac{|J|}{\sqrt{2(K^2+2J^2)}\sqrt{1-\frac{K}{\sqrt{K^2+2J^2}}}}(|2,0\rangle + |0,2\rangle) + \text{sign}(J)\frac{1}{\sqrt{2}}\sqrt{1-\frac{K}{\sqrt{K^2+2J^2}}}|1,1\rangle$ | $2\Delta - K - \sqrt{K^2 + 2J^2}$ | $2\omega - K - \sqrt{K^2 + 2J^2}$ |
| $\frac{|J|}{\sqrt{2(K^2+2J^2)}\sqrt{1+\frac{K}{\sqrt{K^2+2J^2}}}}(|2,0\rangle + |0,2\rangle) - \text{sign}(J)\frac{1}{\sqrt{2}}\sqrt{1+\frac{K}{\sqrt{K^2+2J^2}}}|1,1\rangle$ | $2\Delta - K + \sqrt{K^2 + 2J^2}$ | $2\omega - K + \sqrt{K^2 + 2J^2}$ |

As long as the Kerr-nonlinearity and the coupling are sufficiently weak such that $\omega \pm J > 2K$ and $\omega - K - \sqrt{K^2 + 2J^2}$, then the single-photon states have higher energy than the two-photon states.

In the opposite limit of large two-photon drive strength, $\varepsilon \gg \sqrt{\frac{K}{2}|\Delta \pm J|}$, we can treat the terms $\sim J$, $\Delta$ perturbatively. To leading order, the eigenstates and eigenenergies are given by the two mode cat states listed in the following table:

| Eigenvector | Eigenvalue |
|---|---|
| $|\psi_+\rangle = \frac{1}{\mathcal{N}_+}(|\alpha,\alpha\rangle + |-\alpha,-\alpha\rangle)$ | $\frac{\mathcal{N}_-^2}{\mathcal{N}_+^2}(\Delta + J)|\alpha|^2 = \tanh\left(\frac{2\varepsilon}{K}\right)(\Delta + J)\frac{\varepsilon}{K}$ |
| $|\psi_-\rangle = \frac{1}{\mathcal{N}_-}(|\alpha,\alpha\rangle - |-\alpha,-\alpha\rangle)$ | $\frac{\mathcal{N}_+^2}{\mathcal{N}_-^2}(\Delta + J)|\alpha|^2 = \coth\left(\frac{2\varepsilon}{K}\right)(\Delta + J)\frac{\varepsilon}{K}$ |
| $|\phi_+\rangle = \frac{1}{\mathcal{N}_+}(|\alpha,-\alpha\rangle + |-\alpha,\alpha\rangle)$ | $\frac{\mathcal{N}_-^2}{\mathcal{N}_+^2}(\Delta - J)|\alpha|^2 = \tanh\left(\frac{2\varepsilon}{K}\right)(\Delta - J)\frac{\varepsilon}{K}$ |
| $|\phi_-\rangle = \frac{1}{\mathcal{N}_-}(|\alpha,-\alpha\rangle - |-\alpha,\alpha\rangle)$ | $\frac{\mathcal{N}_+^2}{\mathcal{N}_-^2}(\Delta - J)|\alpha|^2 = \coth\left(\frac{2\varepsilon}{K}\right)(\Delta - J)\frac{\varepsilon}{K}$ |

Here the amplitude is $\alpha \simeq \sqrt{\varepsilon/K}$ and the cat-state normalization factors are $\mathcal{N}_\pm = \sqrt{2}\sqrt{1 \pm e^{-4|\alpha|^2}}$. Note that the states $|\phi_+\rangle$ and $|\psi_+\rangle$ have even total photon number parity, i.e. they are eigenstates of the parity operator $\mathbf{\Pi}_2 = \exp\left[-i\pi\left(\boldsymbol{a}^\dagger \boldsymbol{a} + \boldsymbol{b}^\dagger \boldsymbol{b}\right)\right]$, with eigenvalue $+1$. The states $|\psi_-\rangle$ and $|\phi_-\rangle$ have odd parity.

For $J > 0$ ($J < 0$), the state $|\psi_+\rangle$ ($|\phi_+\rangle$) is the highest energy state (i.e. the roof state). Furthermore, since two-photon drive conserves the total photon number parity, it is clear that the vacuum state at $\varepsilon = 0$ is adiabatically connected with an even photon number parity state at large $\varepsilon$: For $J > 0$ with $|\psi_+\rangle$ and for $J < 0$ with $|\phi_+\rangle$. Similarly, the two excited states with one photon when $\varepsilon = 0$ are adiabatically connected with the odd photon number parity states $|\psi_-\rangle$ and $|\phi_-\rangle$. Schematically:

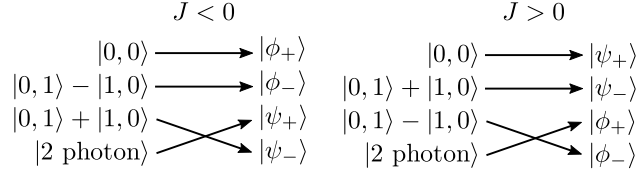

Fig. 2 shows the numerically computed highest part of the intantaneous eigenspectrum of (49). The parameter values used are given in the caption. Note that for $\varepsilon > 0$, this spectrum does not reflect the energy of the system. Indeed, for $\varepsilon \neq 0$, the system is driven and there are no stationary energy eigenstates. Nevertheless, this figure is useful because the adiabatic evolution proceeds along the curves in this spectrum. Interestingly, we find generically, that there are points of crossing (diabolical points[5]), between excited states of different total photon number parity at certain values of the two-photon drive strength.

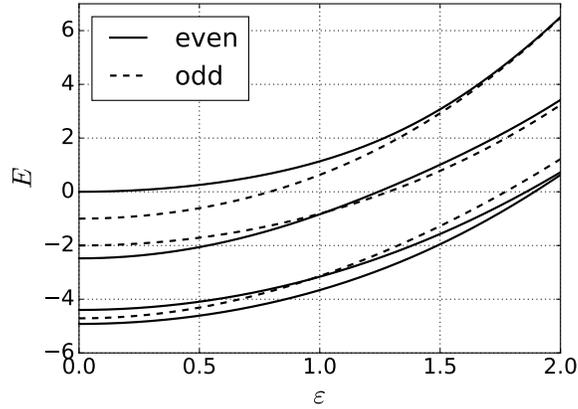

FIG. 2. Highest part of the eigenspectrum of two coupled parametric oscillators in the rotating frame. Solid lines correspond to states with even total photon number parity, while odd lines correspond to states with odd total photon number parity. Notice the crossing points in the spectrum between lines corresponding to excited states with different total photon number parity. These are examples of so-called diabolical points[5]. The parameters used are: $\Delta/\hbar = -1.5\,\mathrm{MHz}$, $K/\hbar = 0.7$, $T \in (0.1, 400)\,\mu s$ and $\varepsilon_{\mathrm{MAX}}/\hbar = 2\,\mathrm{MHz}$.

*1. Diabatic transitions*

In order to transfer the population from the vacuum at $\varepsilon = 0$ to the ground state at $\varepsilon \gg \sqrt{\frac{K}{2}|\Delta \pm J|}$, the drive strength must be increased sufficiently slowly such as to guarantee the adiabatic condition. Here we present numerical results on the breakdown of adiabaticity for two anti-ferromagnetically coupled KPOs. We focus on the case of anti-ferromagnetic coupling $J < 0$. Fig. 3 shows, with the same parameters as Fig. 2, the populations of the highest seven energy instantaneous eigenstates at the end of the ramp, i.e. at $\varepsilon(T) = \varepsilon_{\mathrm{MAX}}$. The populations are shown as a function of increasing normalized linear ramp rate $1/T$, for a system initially in the highest energy state. Clearly, above a certain ramp rate, diabatic transitions from the ground state to other states *with the same total photon number parity*

take place. This can be inferred by comparing Fig. 3 with the spectrum values in Fig. 2 at $\varepsilon = \varepsilon_{\text{MAX}}$. This selection rule reflects the conservation of total photon number parity.

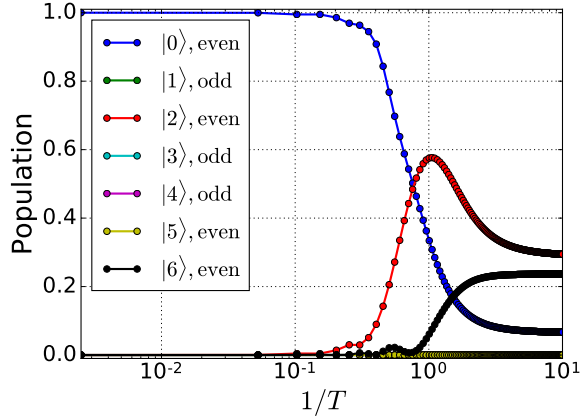

FIG. 3. Populations of the seven highest eigenstates as a function of the normalized ramp rate. The states are labeled as $|n\rangle$, with $n = 0$ denoting the highest eigenvalue. Comparing with Fig. 2, we see that population is transferred out of the highest state into states with the same even total photon number parity. The parameters used in the simulation are: $\Delta/\hbar = -1.5\,\text{MHz}$, $K/\hbar = 0.7$, $T \in (0.1, 400)\,\mu\text{s}$ and $\varepsilon_{\text{MAX}}/\hbar = 2\,\text{MHz}$.

Moving beyond the coherent limit, we next discuss the dissipative case semi-classically. One difference that is apparent from the above discussion, is that single photon loss breaks photon number parity conservation. Hence dissipative transitions between states with different total photon number parity become possible.

### B. Semiclassical analysis

Here we present the key elements of the semi-classical analysis used in the manuscript to convey the intuition about the dissipative optimization process in the quantum Ising machine.

Our starting point are the semi-classical mean-field equations for two coupled oscillators, in a frame rotating with half the drive frequency $\Omega/2$. Formally they can be obtained by replacing the operators in the Heisenberg-Langevin equation with complex functions as $a \to \alpha(t)$ and $b \to \beta(t)$ and averaging out the noise terms. They are (see Methods in the main text)

$$i\dot{\alpha} = (\Delta - 2K|\alpha|^2)\alpha + J\beta + 2\varepsilon\alpha^* - i\frac{\kappa}{2}\alpha, \tag{50}$$

$$i\dot{\beta} = (\Delta - 2K|\beta|^2)\beta + J\alpha + 2\varepsilon\beta^* - i\frac{\kappa}{2}\beta. \tag{51}$$

#### 1. Steady states

In the steady state we have

$$(\Delta - 2K|\alpha|^2)\alpha + J\beta + 2\varepsilon\alpha^* - i\frac{\kappa}{2}\alpha = 0, \tag{52}$$

$$(\Delta - 2K|\beta|^2)\beta + J\alpha + 2\varepsilon\beta^* - i\frac{\kappa}{2}\beta = 0. \tag{53}$$

The three solutions of interest are (i): $\alpha = \beta = 0$, which corresponds to the vacuum state, (ii): $\alpha = \beta = \alpha_s^{(\text{ii})}$, which corresponds to symmetric mode where both oscillators oscillate in phase and (iii): $\alpha = -\beta = \alpha_s^{(\text{iii})}$, which corresponds to the anti-symmetric mode where both oscillators oscillate out-of-phase by $\pi$. As explained in III A, we require that

$\Delta \pm J < 0$, such that the vacuum is the roof state. This implies that solution (ii) exists only if $\varepsilon > \frac{1}{2}\sqrt{(\Delta+J)^2 + \left(\frac{\kappa}{2}\right)^2}$, and the amplitude is then

$$\alpha_s^{(ii)} = \pm e^{i\phi}\sqrt{\frac{\sqrt{4\varepsilon^2 - \left(\frac{\kappa}{2}\right)^2} + \Delta + J}{2K}}. \tag{54}$$

Similarly, solution (iii) exists if $\varepsilon > \frac{1}{2}\sqrt{(\Delta-J)^2 + \left(\frac{\kappa}{2}\right)^2}$, and the amplitude is then

$$\alpha_s^{(iii)} = \pm e^{i\phi}\sqrt{\frac{\sqrt{4\varepsilon^2 - \left(\frac{\kappa}{2}\right)^2} + \Delta - J}{2K}}. \tag{55}$$

Here $\phi = -\frac{1}{2}\arctan\left[\frac{\kappa}{\sqrt{16\varepsilon^2 - \kappa^2}}\right]$.

### 2. Linear stability

Next, we analyze the linear stability of the above solutions. To this end, we substitute $\alpha(t) = \alpha_s + \Delta\alpha(t)$ and $\beta(t) = \beta_s + \Delta\beta(t)$ into (68) and linearize in $\Delta\alpha$ and $\Delta\beta$ to obtain in matrix notation

$$\frac{d}{dt}\begin{pmatrix}\Delta\alpha \\ \Delta\alpha^* \\ \Delta\beta \\ \Delta\beta^*\end{pmatrix} = \begin{pmatrix} -i\Delta + 4iK|\alpha_s|^2 - \frac{\kappa}{2} & 2i(K\alpha_s^2 - \varepsilon) & -iJ & 0 \\ -2i(K\alpha_s^{*2} - \varepsilon) & i\Delta - 4iK|\alpha_s|^2 - \frac{\kappa}{2} & 0 & iJ \\ -iJ & 0 & -i\Delta + 4iK|\beta_s|^2 - \frac{\kappa}{2} & 2i(K\beta_s^2 - \varepsilon) \\ 0 & iJ & -2i(K\beta_s^{*2} - \varepsilon) & i\Delta - 4iK|\beta_s|^2 - \frac{\kappa}{2} \end{pmatrix}\begin{pmatrix}\Delta\alpha \\ \Delta\alpha^* \\ \Delta\beta \\ \Delta\beta^*\end{pmatrix}. \tag{56}$$

We start with the solution (i), $\alpha_s = \beta_s = 0$. In this case the stability equation is

$$\frac{d}{dt}\begin{pmatrix}\Delta\alpha \\ \Delta\alpha^* \\ \Delta\beta \\ \Delta\beta^*\end{pmatrix} = \begin{pmatrix} -i\Delta - \frac{\kappa}{2} & -2i\varepsilon & -iJ & 0 \\ 2i\varepsilon & i\Delta - \frac{\kappa}{2} & 0 & iJ \\ -iJ & 0 & -i\Delta - \frac{\kappa}{2} & -2i\varepsilon \\ 0 & iJ & 2i\varepsilon & i\Delta - \frac{\kappa}{2} \end{pmatrix}\begin{pmatrix}\Delta\alpha \\ \Delta\alpha^* \\ \Delta\beta \\ \Delta\beta^*\end{pmatrix}. \tag{57}$$

The eigenvalues are

$$\lambda_{0,\pm}^{(i)} = -\frac{\kappa}{2} \pm \sqrt{4\varepsilon^2 - (\Delta+J)^2}, \tag{58}$$

$$\lambda_{1,\pm}^{(i)} = -\frac{\kappa}{2} \pm \sqrt{4\varepsilon^2 - (\Delta-J)^2}. \tag{59}$$

The stability of this solution requires that $\text{Re}[\lambda_{0,\pm}], \text{Re}[\lambda_{1,\pm}] < 0$, which in turn requires that

$$\varepsilon < \frac{1}{2}\sqrt{(\Delta \pm J)^2 + \left(\frac{\kappa}{2}\right)^2}. \tag{60}$$

Next we consider solution (ii), $\alpha_s = \beta_s = \alpha_s^{(ii)}$. The Hermitian transformation matrix (56) in this case has the same structure as the one in (57) with the replacements

$$\varepsilon \to \varepsilon - K\alpha_s^2, \tag{61}$$

$$\Delta \to \Delta - 4K|\alpha_s|^2. \tag{62}$$

Hence we can immediately write down the eigenvalues

$$\lambda_{0,\pm}^{(ii)} = -\frac{\kappa}{2} \pm \sqrt{4|\varepsilon - K\alpha_s^2|^2 - (\Delta - 4K|\alpha_s|^2 + J)^2}, \tag{63}$$

$$\lambda_{1,\pm}^{(ii)} = -\frac{\kappa}{2} \pm \sqrt{4|\varepsilon - K\alpha_s^2|^2 - (\Delta - 4K|\alpha_s|^2 - J)^2}. \tag{64}$$

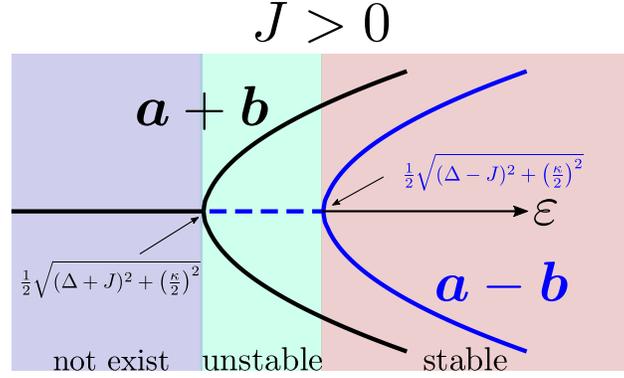

FIG. 4. Stability diagram for the symmetric mode when $J > 0$. Note that we consider the case $\Delta \pm J < 0$ throughout.

After some algebra the condition that the real parts of these eigenvalues are smaller than zero leads to the inequality

$$X + (\Delta \pm J)\sqrt{X} > 0, \tag{65}$$

with $X = 4\varepsilon^2 - \left(\frac{\kappa}{2}\right)^2$. This in turn implies that the solution is stable if and only if

$$\varepsilon > \frac{1}{2}\sqrt{(\Delta \pm J)^2 + \left(\frac{\kappa}{2}\right)^2}. \tag{66}$$

It is easy to see that solution (iii) must satisfy the same stability criterion (66) since the amplitudes only appear squared in the transformation matrix (56). Fig. 4 summarizes these results for the bifurcation behavior of the symmetric mode when $J > 0$. The situation with $J < 0$ is similar but with the roles of symmetric and anti-symmetric modes exchanged.

### C. Entanglement

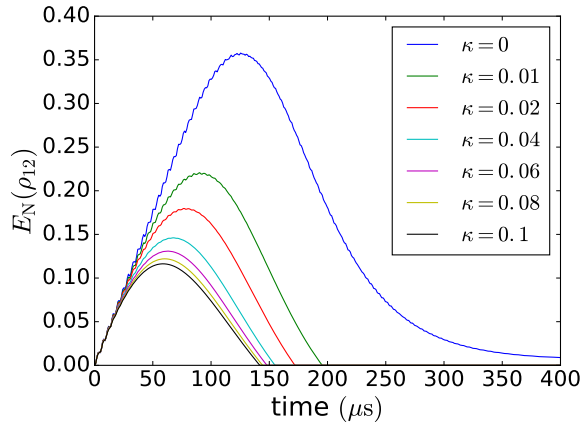

FIG. 5. Logarithmic negativity of the two-mode Gaussian state during the annealing. The covariance matrices are obtained by numerically solving the master equation (Eq. (4) of the main text). A linear ramp is used $\varepsilon(t) = \varepsilon_{\text{MAX}}(t/T)$. The parameters are: $\Delta/\hbar = -1.0\,\text{MHz}$, $K/\hbar = 0.7\,\text{MHz}$, $\varepsilon_{\text{MAX}}/\hbar = 2.0\,\text{MHz}$ and $T = 400\,\mu\text{s}$.

We numerically characterize the presence of entanglement during the annealing by computing the logarithmic negativity

$$E_N(\rho_{12}) = \max(0, -\log_2(2\nu_s)), \tag{67}$$

where $\nu_s$ is the smallest symplectic eigenvalues of the partially transposed Wigner covariance matrix $V$ of the two-mode Gaussian state[6]: $V_{ij} = (R_{ij}+R_{ji})/2$ where $R_{ij} = \langle R_i R_j\rangle - \langle R_i\rangle \langle R_j\rangle$ and $R = (q_1, p_1, q_2, p_2)$ where $q_i = (a_i + a_i^\dagger)/\sqrt{2}$ and $p_i = -i(a_i - a_i^\dagger)/\sqrt{2}$. The logarithmic negativity is an entanglement monotone. If $E_N = 0$ the state is separable. $E_N > 0$ indicates the presence of bipartite entanglement. Fig. 5 shows the logarithmic negativity of the state of the two coupled oscillators during the annealing for different values of the dissipation rate when starting in the two-mode vacuum. Interestingly we find that while dissipation tends to suppress entanglement, transient entanglement is present even with a finite dissipation rate.

### D. Robustness to photon dephasing

In the main text we show that the continuous variable Ising machine (CVIM) displays robustness to photon loss. Here we show that it also provides protection against photon dephasing. Photon dephasing is typically not the dominant decoherence channel[7,8], but can arise for example as a consequence of cross-Kerr terms coupling different oscillators, which have been neglected in I D. Photon dephasing adds additional terms to the Lindblad term in the master equation, which now reads

$$\dot{\rho} = -i[H, \rho] + \kappa \sum_{n=1}^{N} \mathcal{D}[a_n]\rho + \kappa_\phi \sum_{n=1}^{N} \mathcal{D}[a_n^\dagger a_n]\rho. \tag{68}$$

Here $H$ is given by (34) in the rotating frame of the drive. In the following we generalize the arguments of Refs.[9,10]. The master equation (68) can be written equivalently as

$$\dot{\rho} = -i\left(H_{\text{NH}}\rho - \rho H_{\text{NH}}^\dagger\right) + \kappa \sum_{n=1}^{N} a_n \rho a_n + \kappa_\phi \sum_{n=1}^{N} a_n^\dagger a_n \rho a_n^\dagger a_n, \tag{69}$$

with the non-hermitian Hamiltonian

$$H_{\text{NH}} = H - i\frac{\kappa}{2}\sum_{n=1}^{N} a_n^\dagger a_n - i\frac{\kappa_\phi}{2}\sum_n a_n^\dagger a_n a_n^\dagger a_n \tag{70}$$

$$= H - i\frac{\kappa + \kappa_\phi}{2}\sum_{n=1}^{N} a_n^\dagger a_n - i\frac{\kappa_\phi}{2}\sum_n a_n^\dagger a_n^\dagger a_n a_n \tag{71}$$

Hence, we see that photon dephasing results in a renormalization of the single photon loss rate: $\kappa \to \kappa + \kappa_\phi$ and in an imaginary shift of the self-Kerr term: $K \to K - i\frac{\kappa_\phi}{2}$. This will only modify the steady state value of the amplitudes but not the form of the steady state which is still a multi-mode cat state. In addition, the jump part induces random and independent rotations of the *amplitudes*. The transition rate from a configuration $\{s_1, s_2, \ldots, s_n, \ldots, s_N\}$ to a configuration $\{s_1, s_2, \ldots, -s_n, \ldots, s_N\}$ is determined by the transition rate from a coherent state $|\alpha\rangle$ to the state $|-\alpha\rangle$. The latter is estimated as follows. After a jump event the state is

$$\frac{n|\alpha\rangle}{\sqrt{\langle\alpha|n^2|\alpha\rangle}} = \frac{n|\alpha\rangle}{\sqrt{\bar{n}(\bar{n}+1)}} \simeq \frac{n|\alpha\rangle}{\bar{n}}. \tag{72}$$

Here $\bar{n} = |\alpha|^2$ and in the last step we have assumed $\bar{n} \gg 1$. In this limit, the probability that this state is $|-\alpha\rangle$ is

$$P_{|-\alpha\rangle} = \frac{1}{\bar{n}^2}|\langle -\alpha|n|\alpha\rangle|^2 = e^{-4|\alpha|^2}. \tag{73}$$

Furthermore, the probability that in a small time interval $\Delta t$ a jump happens is

$$p_{\text{jump}} = \kappa_\phi \Delta t \langle\alpha|n^2|\alpha\rangle \simeq \kappa_\phi \Delta t \bar{n}^2. \tag{74}$$

Therefore, the rate at which transitions between $|\alpha\rangle$ and $|-\alpha\rangle$ occur due to dephasing is

$$\Gamma_\phi = \frac{p_{\text{jump}} P_{|-\alpha\rangle}}{\Delta t} = \kappa_\phi \bar{n}^2 e^{-4|\alpha|^2}. \tag{75}$$

Hence at large amplitudes (i.e. far above threshold), the adverse effect of photon dephasing on quantum optimization with coupled KPOs is exponentially suppressed in the mean photon number of the coherent states encoding the Ising spins.

### E. Quantum trajectories comparison between CVIM and QA

In Fig. 2 (b), of the main text we show a typical quantum trajectory of the quantum Ising machine with two oscillators in the adiabatic limit and in the presence of photon loss with rate $\kappa = 0.01$. Fig. 6 shows the corresponding behavior for a discrete spin-based quantum annealer with two anti-ferromagnetically coupled qubits. We use a linear ramp of the form $\varepsilon(t) = \varepsilon_{\mathrm{MAX}}(1 - t/T)$, controlling the transverse field part of the quantum Ising model (see Eq. (1) of the main text). The left panel of Fig. 6 shows the coherent case for reference. Here we see that the optimization succeeds and the system evolves from the ground state of the transverse fields $|X_-\rangle|X_-\rangle = \frac{1}{2}(|\uparrow\rangle - |\downarrow\rangle) \otimes (|\uparrow\rangle - |\downarrow\rangle)$, to the ground state of the anti-ferromagnetic Ising model $|\phi_+\rangle = (|\uparrow,\downarrow\rangle + |\downarrow,\uparrow\rangle)/\sqrt{2}$. The additional states shown are $|\psi_\pm\rangle = (|\uparrow,\uparrow\rangle \pm |\downarrow,\downarrow\rangle)$ and $|\phi_-\rangle = (|\uparrow,\downarrow\rangle - |\downarrow,\uparrow\rangle)$. Note that we use the same notation as for the corresponding states of the CVIM in order to emphasize the analogy. The nature of these two sets of states is however different. While the states of the qubit based quantum annealer are eigenstates of an operator with discrete spectrum, living in a finite Hilbert space, the corresponding states of the quantum Ising machine are eigenstates of an operator with a continuous spectrum and the associated Hilbert space is infinite dimensional. The right panel of Fig. 6 shows a typical quantum trajectory for the same parameters, when the system is subject to dephasing errors with rate $\gamma/\hbar = 0.01$ MHz, in the basis that diagonalizes the Ising term, i.e. the jump operators are $\sigma_1^z$ and $\sigma_2^z$. Clearly, we see that dephasing errors quickly lead to large fluctuations in the populations of the states with correct Ising encoding ($|\phi_\pm\rangle$). In the example shown, the state at the end of the adiabatic ramp is clearly not the ground state of the anti-ferromagnetic Ising model.

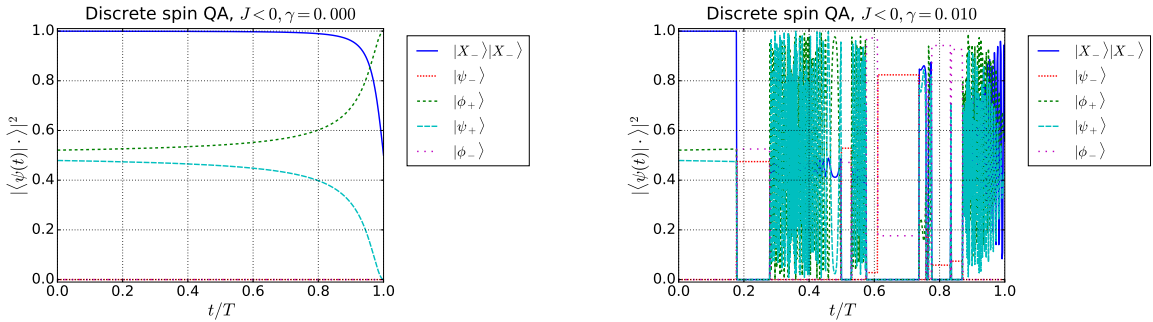

FIG. 6. Effect of dephasing on the discrete qubit quantum annealer. **Left panel**: coherent limit $\gamma = 0$. Here the annealing succeeds and the state at the end of the ramp is the ground state of the anti-ferromagnetic Ising model $|\phi_+\rangle = (|\uparrow,\downarrow\rangle + |\downarrow,\uparrow\rangle)/\sqrt{2}$. **Right panel**: Incoherent case with dephasing rate $\gamma/\hbar = 0.01\{\mathrm{MHz}\}$

## IV. TYPICAL QUANTUM TRAJECTORIES FOR THE NUMBER PARTITIONING PROBLEM

In Fig. 4 (a) of the main text, we show the success probability of the quantum Ising machine for the number partitioning with the set $S = \{4, 5, 6, 7\}$. Figs. 7, 8 and 9 show the inter-oscillator phase-correlations $\langle a_n^\dagger a_m\rangle$, along typical quantum trajectories corresponding to different points in Fig. 4 (a) in the main text. For two elements $n$ and $m$ in the same partition, the correlation is ideally $\pi$ while it is zero if they are in different partitions. Explicitly, the correct solution to the considered partitioning problem has the following correlations above threshold

$$|\arg(\langle a_1^\dagger a_2\rangle)| = |\arg(\langle a_1^\dagger a_3\rangle)| = |\arg(\langle a_2^\dagger a_4\rangle)| = |\arg(\langle a_3^\dagger a_4\rangle)| = \pi \tag{76}$$

$$|\arg(\langle a_2^\dagger a_3\rangle)| = |\arg(\langle a_1^\dagger a_4\rangle)| = 0. \tag{77}$$

Note that the phases only have meaning when $\langle a_n^\dagger a_m\rangle \ne 0$. Clearly, we see by comparing Figs. 7, 8 and 9 with Fig. 4 (a) of the main text, that for parameter values corresponding to the blue region in Fig. 4 (a), the typical trajectories display the correct correlations, while for parameters in the red regions, the correlations do typically not all have the correct values at the end of the ramp. We also see that in regions with higher number of photon loss events, the noise in the conditional expectation values increases. For example, the trajectories at $\kappa = 0$ and $\kappa = 0.04$ for $\dot\varepsilon = 3$, in Fig. 9 are similar to each other. This corresponds to the regime of low average number of photon jumps. In contrast, the

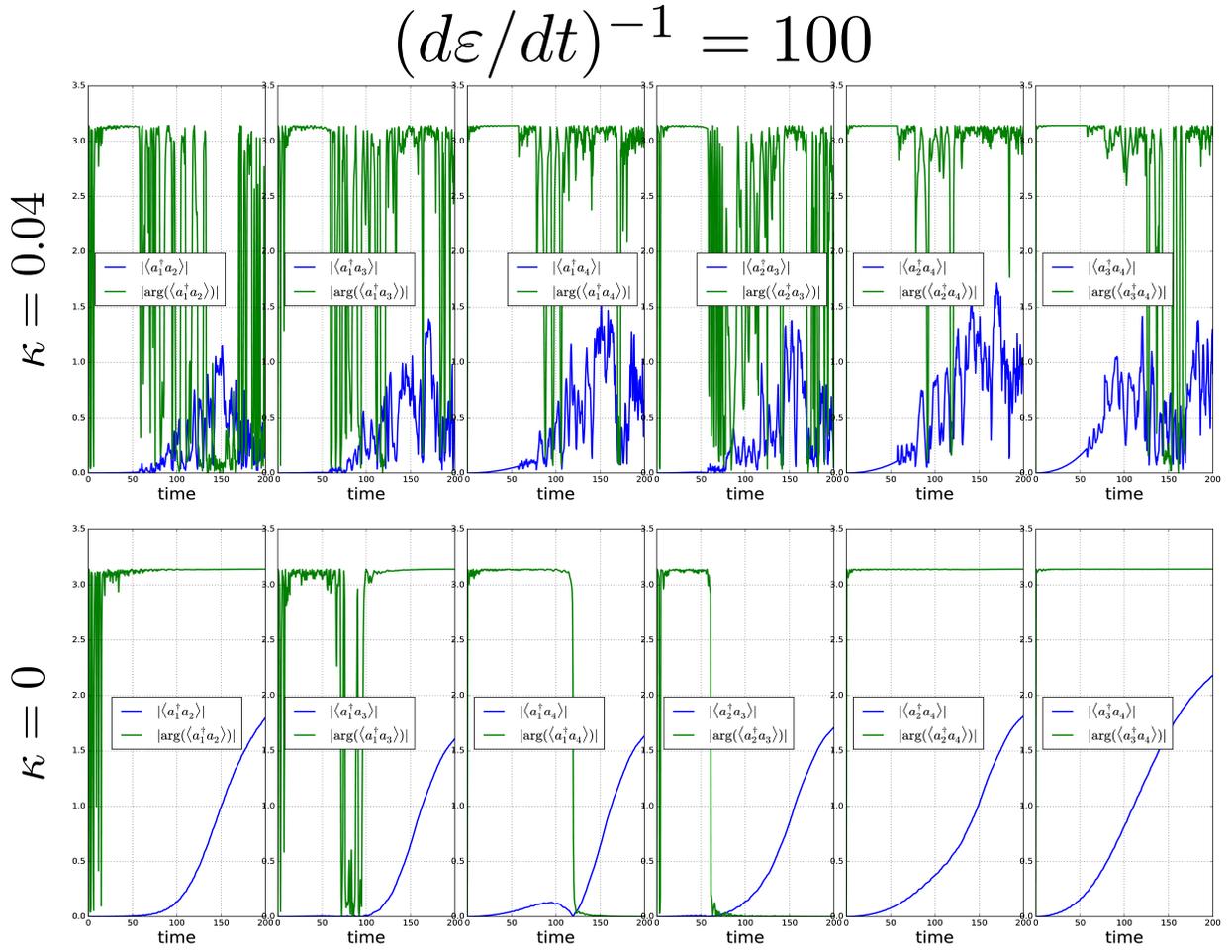

FIG. 7. Conditional expectation values of the inter-oscillator phase correlations for $\dot{\varepsilon} = 100$

trajectories with $\kappa = 0$ and $\kappa = 0.04$ for $\dot{\varepsilon} = 100$, in Fig. 7, are markedly different. This corresponds to a regime with a higher average number of photon jumps (see Fig.4 (b) in the main text).

$$(d\varepsilon/dt)^{-1} = 15$$

FIG. 8. Conditional expectation values of the inter-oscillator phase correlations for $\dot{\varepsilon} = 15$

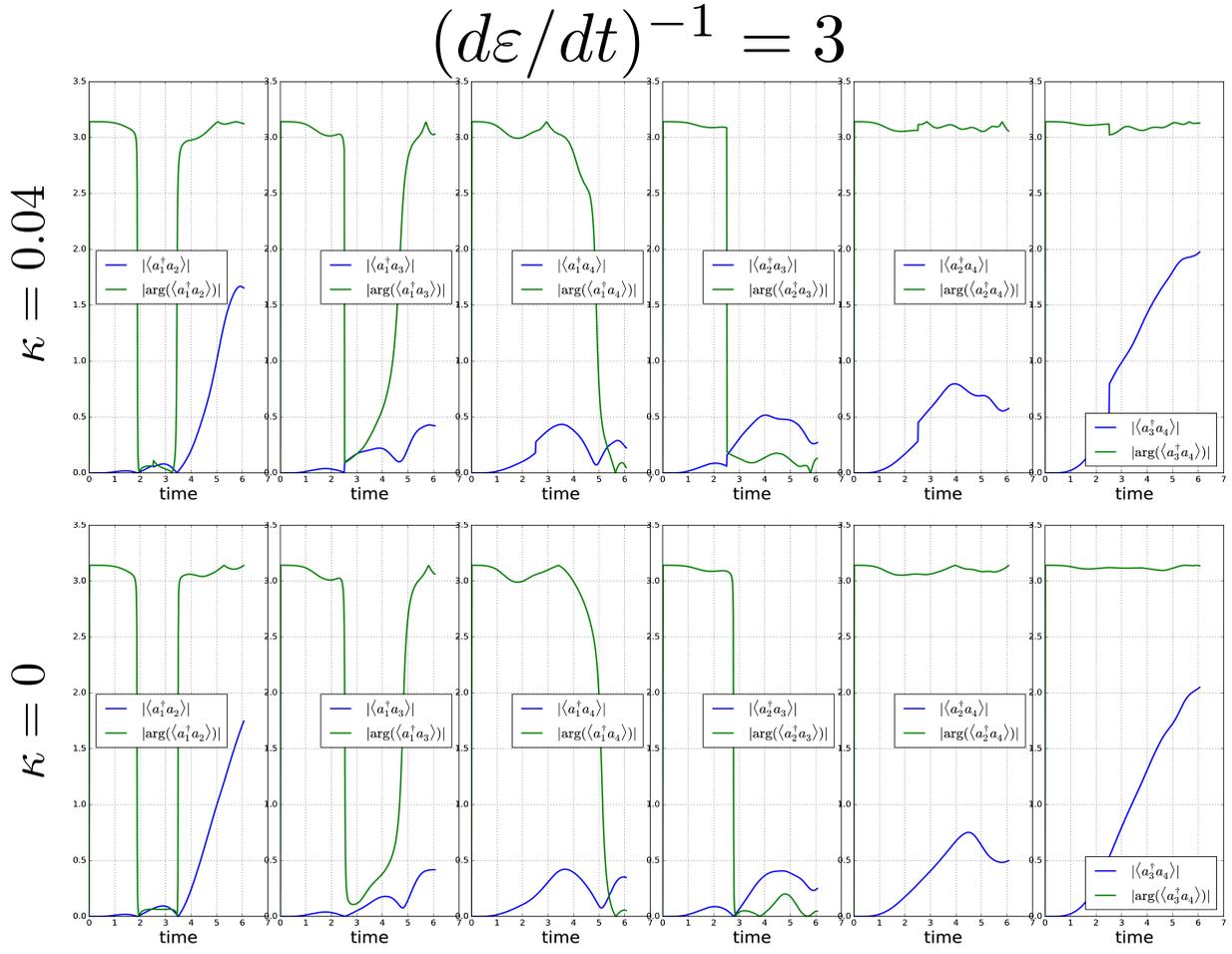

FIG. 9. Conditional expectation values of the inter-oscillator phase correlations for $\dot{\varepsilon} = 3$



\end{document}